\documentclass[final,1p,times,authoryear]{elsarticle}

\usepackage{amssymb}
\usepackage{lipsum}
\usepackage{bm}
\usepackage{amsmath}

\journal{New Astronomy}

\begin{document}

\begin{frontmatter}

\title{Modifications of SPH towards three-dimensional simulations of an icy moon with internal ocean}

\author[label1]{Keiya Murashima}
\affiliation[label1]{organization={Department of Astronomy, Kyoto University},
             addressline={Kitashirakawa-Oiwake-cho},
             city={Sakyo-ku},
             postcode={606-8502},
             state={Kyoto},
             country={Japan}}

\author[label2]{Natsuki Hosono}
\affiliation[label2]{organization={Center for Planetary Science Integrated Research Center of Kobe University},
             addressline={7-1-48 Minatojima-Minamimach},
             city={Chuo-ku},
             postcode={650-0047},
             state={Kobe},
             country={Japan}}

\author[label3]{Takayuki R. Saitoh}
\affiliation[label3]{organization={Department of Planetology, Kobe University},
             addressline={1-1 Rokkodai-cho},
             city={Nada-ku},
             postcode={657-8501},
             state={Kobe},
             country={Japan}}

\author[label1]{Takanori Sasaki}

\begin{abstract}
There are some traces of the existence of internal ocean in some icy moons, such as the vapor plumes of Europa and Enceladus. 
This implies a region of liquid water beneath the surface ice shell. Since liquid water would be essential for the origin of life, it is important to understand the development of these internal oceans, particularly their temperature distribution and evolution. 
The balance between tidal heating and radiative cooling is believed to sustain liquid water beneath an icy moon's surface. 
We aim to simulate the tidal heating of an internal ocean in an icy moon using 3-dimensional numerical fluid calculations with the Smoothed Particle Hydrodynamics (SPH) method. 
We incorporated viscosity and thermal conduction terms into the governing equations of SPH. 
However, we encountered two issues while calculating rigid body rotation using SPH with a viscous term: (1) conventional viscosity formulations generated unphysical forces that hindered rotation, and (2) there was artificial internal energy partitioning within the layered structure, which was due to the standard SPH formulations. 
To address the first issue, we modified the viscosity formulation.
For the second, we adopted Density Independent SPH (DISPH) developed in previous studies to improve behavior at discontinuous surfaces.
Additionally, we implemented radiative cooling using an algorithm to define fluid surfaces via the particle method. 
We also introduced an equation of state accounting for phase transitions. 
With these modifications, we have refined the SPH method to encompass all necessary physical processes for simulating the evolution of icy moons with internal oceans.
\end{abstract}



\begin{keyword}
Hydrodynamical simulations \sep Icy satellites



\end{keyword}

\end{frontmatter}




\section{Introduction}
\label{introduction}

``Does extraterrestrial life exist?'' is one of the most important questions in science.
In order to address this question, there has been a lot of discussions about the habitability of life and the exploration of planets that may be able to host life.
The internal oceans of icy moons have emerged as promising candidates for harboring extraterrestrial life.
Observations indicate that water vapor is being emitted from cracks in the ice shells of some icy moons, such as Europa and Enceladus \citep[e.g.,][]{porco2006, sparks2016}.
This suggests a subsurface region of liquid water.
Icy moons, both within and beyond our solar system, are hypothesized to be widespread celestial bodies harboring liquid water. 
Since liquid water is fundamental for chemical reactions and is considered crucial for life's origin, understanding the development of these internal oceans, particularly their temperature distribution and evolution, is essential.

The thermal evolution of icy moons' internal oceans is thought to involve: (1) Tidal heating in the ice shell due to tidal deformation, most commonly, the warm ice at the base is the most dissipative,
(2) transfer of heat to the satellite's surface via thermal conduction, and (3) loss of heat through surface radiation.
The equilibrium of these processes maintains the existence of the internal ocean.
Additionally, a phase transition between ice and water occurs at the boundary between the ice shell and the interior ocean.
Consequently, in order to calculate the structure and evolution of the internal ocean, it is necessary to consider at least the following four physical processes: tidal heating, heat conduction, radiative cooling, and phase transition.
Generally, tidal deformation is attributed to three factors: (1) Changes in the satellite bulge size due to its elliptical orbit, (2) longitudinal shift of this bulge, and (3) latitudinal bulge shift due to the rotation axis's inclination.
Therefore, three-dimensional global simulations are required to solve for tidal heating due to the three-dimensional tidal deformation induced by the above factors.

Previous studies have numerically calculated the thermal evolution of internal oceans \citep[e.g.,][]{tobie2003, ashkenazy2018}, focusing either on parts of the ice shell or employing two-dimensional models for the entire satellite.

Our objective was to perform a comprehensive three-dimensional simulation of an entire icy moon using the Smoothed Particle Hydrodynamics (SPH) method. We aimed to (1) determine the necessary conditions for the formation and maintenance of an internal ocean by varying parameters such as satellite mass, internal structure, and orbital distance, and (2) elucidate the current internal structure and heat dynamics of solar system icy moons using our developed simulation model.

SPH method, developed for astrophysical phenomena simulation by \cite{lucy1977} and \cite{gingold1977}, represents fluid as a collection of hypothetical particles (SPH particles).
Each particle possesses physical properties, with governing equations formulated through interactions with neighboring particles.
The SPH method offers several advantages: (1) Ease of programming for three-dimensional calculations due to minimal dimensional dependence, (2) Galilean invariance, (3) simplicity in integrating new physical processes, (4) high resolution in high-density regions, (5) suitability for problems involving significant deformations, coalescence, or destruction, (6) guaranteed conservation of angular momentum in advective terms, and (7) compatibility with parallel GPU computing.
At the boundary between the satellite's ice shell and the interior ocean, solid ice and liquid water seem to be mixed.
Moreover, large-scale parallelized calculations are necessary for parameter studies or high-resolution simulations. 
These advantages make SPH ideal for simulating the internal oceans of icy moons in three dimensions.

This study aims to develop an SPH code capable of simulating all aspects of icy moons' thermal evolution, including tidal heating, heat conduction, and radiative cooling.
However, the conventional SPH method faces several challenges: (1) For tidal heating, a conservative viscosity implementation in SPH leads to unphysical momentum transfer, falsely inhibiting rigid body rotation, (2) an unnatural internal energy distribution arises in layered particle surfaces due to standard SPH formulations, and (3) a lack of sophisticated techniques for radiative cooling to detect surfaces within SPH.
In this paper, we propose prescriptions for these three problems.
For (1), we identified that the ``pairwise'' conservative viscosity implementation disrupts angular momentum conservation, guaranteed by SPH method, although it is an advantage of SPH method to mesh-base schemes.
Thus, we propose a non-pairwise viscosity formulation that preserves momentum.
For (2), the artificial particle structure, problematic in ``standard'' SPH (SSPH), is addressed by employing Density Independent SPH (DISPH) \citep{saitoh2013, hosono2013a, hopkins2013}, which better manages surfaces compared to SSPH.
For (3), we develop a method for implementing radiative cooling from surfaces.
In this paper, we deals only with modifications of the SPH method.
We will demonstrate through simple test simulations how our implementations effectively resolve these issues.

We note that several other methods have been proposed to solve problem (2) \citep[e.g.][etc.]{wadsley2017gasoline2, pearl2022fsisph, yuasa2024}.
One approach to solve this problem is to modify the definition of density.
DISPH is one of the methods that has taken this approach.
\citet{wadsley2017gasoline2} is another method that has taken this approach and, like DISPH, has achieved good performance in the tests of \citet{agertz2007fundamental}.
Another approach is to use a Riemann solver.
Godunov SPH, which was developed by \citet{inutsuka1994godunov, inutsuka2002reformulation}, determines the physical quantities between particles by the Riemann problem and uses them for interaction calculations.
Other approaches using Riemann solvers include Contact SPH \citep{parshikov2000improvements}, FSISPH \citep{pearl2022fsisph} and DIGSPH \citep{yuasa2024} etc.
The approach based on the Riemann solver yields good results in multi-fluid mixing problems, however these methods are semi-implicit and have the limitation of being computationally expensive.
Another approach is to use artificial heat conduction.
This was proposed by \citet{price2008modelling} and gives differentiability to the thermal energy field by introducing artificial diffusion into the energy equation.
However, this method is not applicable to any equation of state other than the equation of stateof ideal gas.

In section \ref{sec:BasicEquations}, we will present the governing equations. Section \ref{sec:SPH} will introduce our SPH formulations for simulating icy moons using SSPH and DISPH. In section \ref{sec:results}, we conduct rigid body rotation tests. Finally, section \ref{sec:conclusion} will summarize the paper.

\section{Basic equations} \label{sec:BasicEquations}
\lipsum[1]

\subsection{Governing equations}
The governing equations for an icy satellite are formulated as follows
\begin{eqnarray}
    &&\frac{d\rho}{dt} = - \rho\nabla\cdot\bm{v},\\
    &&\frac{d\bm{v}}{dt} = \frac{1}{\rho}\nabla p + \frac{1}{\rho}\nabla \cdot \mathbf{\Pi} + \bm{g}, \label{eqmotion}\\
    &&\frac{du}{dt} = - \frac{p}{\rho}\nabla\cdot\bm{v} + \frac{1}{\rho}\mathbf{\Pi}:\mathbf{e} +\nabla\cdot(k\nabla T) - \dot{u}^{\rm rad}, \label{eqenergy}
\end{eqnarray}
where $\rho$ is density, $t$ is time, $\bm{v}$ is the velocity, $p$ is pressure, $\bm{g}$ is gravity, $u$ is the specific internal energy, $k$ is the thermal conductivity, $T$ is temperature, $\dot{u}^{\rm{rad}}$ is the term for radiative cooling, $\mathbf{\Pi}$ is the viscous stress tensor, and $\mathbf{e}$ is the strain rate tensor. In the equation of energy, we consider $u$ as the independent variable, with $p$ and $T$ being calculated using the equation of state (refer to section \ref{subsec:EOS} for details). The expressions for $\mathbf{\Pi}$ and $\mathbf{e}$ are given by
\begin{eqnarray}
    &&\mathbf{\Pi} = \mu\left[(\nabla\otimes\bm{v}) + (\nabla\otimes\bm{v})^{\mathsf T} - \frac{2}{3}(\nabla\cdot\bm{v})\mathbf{I}_3\right], \\
    &&\mathbf{e} = \frac{1}{2}\left[ (\nabla\otimes\bm{v}) + (\nabla\otimes\bm{v})^{\mathsf T}\right],
\end{eqnarray}
where $\mu$ is the coefficient of viscosity, $\mathbf{I}_3$ is the three-dimensional identity matrix, and the superscript $\mathsf{T}$ represents the transposed matrix.

\begin{table}
\begin{center}
\begin{tabular}{lcccccc}
\hline
Material & $a_0$ & $b_0$ & $A_0$ & $B_0$ & $\rho_0$ & $u_0$  \\
& & & (GPa) & (GPa) & ($\rm kg/m^3$) & (MJ/kg) \\
\hline
Ice & 0.3 & 0.1 & 9.47 & 9.47 & 917.0 & 10.0   \\
Water & 0.7 & 0.15 & 2.18 & 2.18 & 998.0 & 7.0   \\ 
Granite & 0.5 & 1.3 & 18.0 & 18.0 & 2680.0 & 16.0  \\\hline
\end{tabular} \label{table:tillotson}
\caption{Material-specific parameters of Tillotson EOS; These parameters are listed in \cite{melosh1989}, p. 234, Table AII.3.}
\end{center}
\end{table}

\subsection{Equation of state}\label{subsec:EOS}

Regarding the EOS for the rocky core, we employ the Tillotson EOS \citep{tillotson1962}, formulated as follows
\begin{equation}
    p = \left(a_0 + \frac{b_0}{\frac{u}{u_0\eta^2}+1}\right)\rho u \,+\,A_0(\eta - 1) \,+\, B_0(\eta -1)^2,
\end{equation}
where $\eta = \rho/\rho_0$ and $a_0, b_0, A_0, B_0, \rho_0, u_0$ are material-specific parameters (refer to table \ref{table:tillotson}). For the rocky core's temperature, we assume a direct proportionality to the internal energy, represented by
\begin{equation}
    u = c_{V} T,
\end{equation}
where $c_{V}$ is the specific heat at constant volume.

In this paper, phase transitions are not considered. 
Consequently, we have adopted Tillotson EOS for icy mantle.
In selecting an equation of state (EOS) for $\rm{H_2 O}$, it's essential to choose one that accounts for the phase transition between ice and water. 
Therefore, for the application to realistic icy moons, we are planning to adopt the AQUA-EOS, developed by \cite{haldemann2020a} for calculating the interiors of planets.
This EOS covers a wide range from $0.1,{\rm Pa}$ to $400,{\rm TPa}$ and temperatures ranging from $150,{\rm K}$ to $10^5,{\rm K}$. 
Utilizing AQUA-EOS allows us to accurately compute the pressure and temperature from the internal energy and density.

\subsection{Viscosity model}

In our model, the viscosity of the ice shell is represented using the Frank–Kamenetskii approximation, a common choice in thermal convection models \citep[e.g.,][]{reese1999a, tobie2003}:
\begin{eqnarray}
    \mu_{\rm ice} = \mu_m\left[-\beta (T - T_m)\right], \label{viscmodel}
\end{eqnarray}
where $\mu_m$ denotes the viscosity at the melting temperature, $T_{m}$. We have set $\mu_m$ to be $10^{13},{\rm Pa.s}$. The coefficient $\beta$ is given by
\begin{eqnarray}
    \beta = \frac{Q}{R T_m^2},
\end{eqnarray}
where $Q$ represents the activation energy, and $R$, equal to $8.31,{\rm m^2.kg.s^{-2}.K^{-1}.mol^{-1}}$, is the gas constant \citep{reese1999a}. With $Q$ set at $50,{\rm kJ.mol^{-1}}$, this results in $\beta = 0.08$. For the viscosity of water and the rock core, we have assigned values of $\mu_{\rm{water}} = 0,{\rm Pa.s}$ and $\mu_{\rm{core}} = 1.0 \times 10^{21},{\rm Pa.s}$, respectively.

\subsection{Thermal conductivity}

The thermal conductivity values for ice Ih and high-pressure ice (ice I\hspace{-1pt}I\hspace{-1pt}I) are determined as per the following equations, referenced from \cite{andersson2005}:
\begin{eqnarray}
    k_{\mathrm{Ih}} &=& \frac{632.0}{T} + 0.38 + 0.00197 T, \\
    k_{\mathrm{hp}} &=& 93.2 \times T^{-0.822}. 
\end{eqnarray}
For water and the rocky core, we have set the thermal conductivity values to 
\begin{eqnarray}
    k_{\mathrm{water}} &=& 0.556\,\,\, {\rm W.m^{-1}.K}, \\
    k_{\mathrm{core}} &=& 3.0\,\,\, {\rm W.m^{-1}.K}
\end{eqnarray}
 \citep{hill1962, kirk198791}, respectively.

\subsection{Radiative cooling}

We model the radiation emitted from the surface of the icy moon based on the principle of Black body radiation. The rate of radiative cooling is expressed as
\begin{eqnarray}
    \dot{u}^{\rm rad} = - \frac{\sigma T_{\rm Surface}^4 S_{\rm Surface}}{m},
\end{eqnarray}
where $\sigma = 5.67 \times 10^{-8},{\rm W.m^{-2}.K^{-4}}$ represents the Stefan–Boltzmann constant. Here, $T_{\rm Surface}$ is the surface temperature, and $S_{\rm Surface}$ denotes the surface area.

\section{Numerical method} \label{sec:SPH}
In this section, we explain the numerical methods we use in our simulations.
First, we describe outlines of formulations for pressure gradient, traditional viscosity and thermal conductivity terms within the frameworks of SSPH and DISPH (\ref{sec::outline}).
And then, we introduce our new formulation of viscosity term in SSPH and DISPH, respectively (\ref{sec::new_visc}).
We show radiative cooling terms in section (\ref{sec:radiative}) and numerical time step in section(\ref{sec:timestep}).

\subsection{Outline of SSPH and DISPH}\label{sec::outline}

\subsubsection{Outline of SSPH}

First, we derive the term contributed by the hydrodynamical force in SSPH. In the SPH method, physical quantities are approximated through convolution with a kernel function $W(\bm{r},h)$, as follows
\begin{eqnarray}
    \langle f(\bm{r})\rangle = \int f(\bm{r}^{\prime})W(|\bm{r} - \bm{r}^{\prime}|;h)d\bm{r}^{\prime}, \label{fequation}
\end{eqnarray}
where brackets means the ``smoothed'' value, $\bm{r}$ is the position vector, and $h$ is the smoothing length, representing the effective reach of the kernel function.

The smoothing length $h$ is evaluated by
\begin{eqnarray}
    h_i = 1.2\left(\frac{m_i}{\rho_i}\right)^{1/D},
\end{eqnarray}
with $D$ being the number of dimensions. The kernel function must satisfy three conditions: (1) compact support, (2) normalization, and (3) it becomes the delta function in the $h\to0$ limit.

We use the Wendland C6 kernel for 3D \citep{wendland1995} as our kernel function:
\begin{eqnarray}
    W(\bm{r};h) = \frac{1365}{64\pi H^3}\times
    \left\{
    \begin{array}{ll}
        (1-s)^8 (1 + 8s + 25s^2 + 32s^3) & \text{for}\, 0 \leq s \leq 1 \\
        0 & \mbox{for}\, s > 1,
    \end{array}
    \right.
\end{eqnarray}
where $H (=ah)$ is the kernel function size and $s = |\bm{r}|/H$. In the Wendland kernel, $a = 2$. This kernel is stable against pairing instability \citep{dehnen2012}.

From equation (\ref{fequation}), the first derivative of the smoothed $f$ is
\begin{eqnarray}
    \langle \nabla f(\bm{r})\rangle 
    &=& \int \nabla f(\bm{r}^{\prime})W(|\bm{r} - \bm{r}^{\prime}|;h)d\bm{r}^{\prime} \nonumber \\
    &=& \int f(\bm{r}^{\prime}) \nabla W(|\bm{r} - \bm{r}^{\prime}|;h)d\bm{r}^{\prime}. \label{fdotequation}
\end{eqnarray}
Here, we utilize the compact support property of the kernel function.

To convert the integral to summation, we replace the volume element $d\bm{r}$ with $\Delta V = m / \rho$, leading to discretized equations:
\begin{eqnarray}
    f_i &=& \sum_j f_j \frac{m_j}{\rho_j} W_{ij}(h_i), \label{eq:discretedf_SSPH}\\
    \nabla f_i &=& \sum_j f_j \frac{m_j}{\rho_j} \nabla W_{ij}(h_i), \label{eq:discretedfdot_SSPH}
\end{eqnarray}
where $W_{ij}(h_i) = W(|\bm{r}_i - \bm{r}_j|, h_i)$.
By substituting $\rho$ into $f$ in equation (\ref{eq:discretedf_SSPH}), we derive
\begin{eqnarray}
    \rho_i = \sum_j m_j W_{ij}(h_i). \label{dens_smth}
\end{eqnarray}

Utilizing equations (\ref{eq:discretedf_SSPH}) and (\ref{eq:discretedfdot_SSPH}), the motion and energy equations are
\begin{eqnarray}
    &&\left(\frac{d\bm{v}_i}{dt}\right)^{\rm hydro} = -\sum_j m_j \left(\frac{p_i}{\rho_i^2}+\frac{p_j}{\rho_j^2}\right)\nabla \tilde{W}_{ij}, \\
    &&\left(\frac{du_i}{dt}\right)^{\rm hydro} = \frac{p_i}{\rho_i^2}\sum_j m_j \bm{v}_{ij} \nabla \tilde{W}_{ij}
\end{eqnarray}
where $\bm{v}_{ij} = \bm{v}_i - \bm{v}_j$.
In order to satisfy the Newton's third low, we replaced $W_{ij}(h_i)$ into $\tilde{W}_{ij} = 0.5[W_{ij}(h_i) + W_{ij}(h_j)]$.

Following \cite{hosono2016b}, we employ the von Neumann-Richtmyer-Landshoff (vNRL) artificial viscosity (AV) \citep{richtmyer1948, vonneumann1950, landshoff1955}. In the vNRL AV, an artificial pressure $p^{AV}$ is added to the pressure, calculated as
\begin{eqnarray}
    p^{\rm AV}_i = 
    \left\{
    \begin{array}{ll}
        - \alpha_i^{\rm AV}\rho_i c_i h_i (\nabla \cdot \bm{v})_i + 2\alpha_i^{\rm AV} \rho h_i^2 (\nabla\cdot\bm{v})^2_i & (\nabla\cdot\bm{v})_i < 0, \\
        0 & (\nabla\cdot\bm{v})_i \geq 0,\\
    \end{array}
    \right.
\end{eqnarray}
where $c$ is the sound speed and $\alpha^{\rm{AV}}$ is the parameter determining the strength of artificial viscosity. We set $\alpha^{\rm{AV}} = 1.0$. The required timestep for this AV is given by
\begin{eqnarray}
    \Delta t_i^{\rm CFL} = C^{\rm CFL}\frac{h_i}{h|\nabla\cdot\bm{v}_i| + c_i + 1.2(\alpha^{\rm AV}_i c_i + 2\alpha_i^{\rm AV} h_i |{\rm min}(\nabla\cdot\bm{v}_i, 0)|)}, \label{timestepCFL}
\end{eqnarray}
where $C^{\rm CFL}$ is the CFL coefficient. To reduce shear viscosity, we employ the Balsara switch \citep{balsara1995}.

In SSPH, the conventional viscosity formulation, as outlined by \cite{morris1997}, expresses the momentum diffusion term as
\begin{eqnarray}
    \left(\frac{d\bm{v}_i}{dt}\right)^{\rm visc} = \sum_j\frac{m_j}{\rho_i \rho_j} \frac{4\mu_i \mu_j}{\mu_i + \mu_j}\frac{{\bm r}_{ij}\cdot\nabla \tilde{W}_{ij}}{{\bm r}^2_{ij}}\bm{v}_{ij}, \label{eq:viscforce_RV}
\end{eqnarray}
where $\bm{r}{ij} = \bm{r}i - \bm{r}j$.
The associated viscous heating term, reflecting the energy transfer from kinetic to internal energy, is given by
\begin{eqnarray}
    \left(\frac{du_i}{dt}\right)^{\rm visc} = - \sum_j\frac{m_j}{\rho_i \rho_j}\frac{2\mu_i \mu_j}{\mu_i + \mu_j}\frac{{\bm r}_{ij}\cdot\nabla \tilde{W}_{ij}}{{\bm r}^2_{ij}}{\bm v}_{ij}^2. \label{eq:vischeat_RV}
\end{eqnarray}

\cite{cleary1999} proposed a method for thermal conduction in standard SPH by utilizing the harmonic mean of conduction coefficients. The thermal conduction term is expressed as
\begin{eqnarray}\label{eq::cond_ssph}
    \left(\frac{du_i}{dt}\right)^{\rm cond} = \sum_j\frac{m_j}{\rho_i \rho_j}\frac{4k_i k_j}{k_i + k_j}\frac{{\bf r}_{ij}\cdot\nabla \tilde{W}_{ij}}{{\bf r}^2_{ij}}(T_i - T_j). \label{conductionSSPH}
\end{eqnarray}

\subsubsection{outline of DISPH}

In the DISPH method, a new quantity:
\begin{eqnarray}
   Y_i = \langle p_i^{\alpha}\rangle \Delta V_i, \label{eq:Ydef}
\end{eqnarray}
is introduced.
The $\alpha$ is a constant value less than unity, introduced to enhance the method's behavior under strong shocks.
In subsequent discussions, we denote $p^{\alpha}$ as $y$. The smoothed value $\langle y\rangle$ is calculated as
\begin{equation}
    \langle y_i \rangle = \sum_j Y_jW_{ij}(h_i). \label{eq:y_smth}
\end{equation}

Following the approach outlined by \cite{saitoh2013} and \cite{hosono2013a}, the equations of motion and energy in DISPH are formulated as
\begin{eqnarray}
    &&\left(\frac{d\bm{v}_i}{dt}\right)^{\sf hydro} = -\sum_j \frac{Y_i Y_j}{m_i}\left[\frac{\langle y_i\rangle^{1/\alpha-2}}{\Omega_i}\nabla W_{ij}(h_i) + \frac{\langle y_j\rangle^{1/\alpha-2}}{\Omega_j}\nabla W_{ij}(h_j)\right], \\
    &&\left(\frac{du_i}{dt}\right)^{\sf hydro} = \sum_j \frac{Y_i Y_j}{m_i}\frac{\langle y_i\rangle^{1/\alpha-2}}{\Omega_i}\bm{v}_{ij}\cdot\nabla W_{ij}(h_i),
\end{eqnarray}
where $\Omega$ is the ``grad-h'' term \citep[e.g.,][]{hopkins2013,hosono2013a}:
\begin{eqnarray}
    \Omega_i = 1 + \frac{h_i}{D \langle \rho_i\rangle}\frac{\partial \langle \rho_i \rangle}{\partial h_i}.
\end{eqnarray}\par

Numerical integration within this framework requires the determination of $\langle y_i\rangle$. Due to the implicit relationship between equation (\ref{eq:y_smth}) and the equation of state, we resolve equation (\ref{eq:y_smth}) iteratively (see \cite{hosono2013a} for detail).

Similar to the SSPH method, we employ the von Neumann-Richtmyer-Landshoff (vNRL) artificial viscosity (AV) in DISPH (see \cite{hosono2016b} for detail).

Following \citet{takeyama2017}, the conventional viscosity formulation in DISPH expresses the momentum diffusion term as
\begin{eqnarray}
    \left(\frac{d\bm{v}_i}{dt}\right)^{\rm visc} = \frac{1}{m_i}\sum_j\frac{Y_i Y_j}{\langle y_i\rangle \langle y_j\rangle} \frac{4\mu_i \mu_j}{\mu_i + \mu_j}\frac{{\bm r}_{ij}\cdot\nabla \tilde{W}_{ij}}{{\bm r}^2_{ij}}\bm{v}_{ij}, \label{eq:viscforce_RV}
\end{eqnarray}
where $\bm{r}{ij} = \bm{r}i - \bm{r}j$.
The associated viscous heating term, reflecting the energy transfer from kinetic to internal energy, is given by
\begin{eqnarray}
    \left(\frac{du_i}{dt}\right)^{\rm visc} = - \frac{1}{m_i}\sum_j\frac{Y_i Y_j}{\langle y_i\rangle \langle y_j\rangle}\frac{2\mu_i \mu_j}{\mu_i + \mu_j}\frac{{\bm r}_{ij}\cdot\nabla \tilde{W}_{ij}}{{\bm r}^2_{ij}}{\bm v}_{ij}^2. \label{eq:vischeat_RV}
\end{eqnarray}

In the DISPH framework, the thermal conduction equation can be adapted from the SSPH formulation.
By substituting $m/\rho$ with $Y/\langle y\rangle$ in the equation (\ref{eq::cond_ssph}), the thermal conduction term in DISPH is reformulated as
\begin{eqnarray}
\left(\frac{du_i}{dt}\right)^{\rm cond} = \frac{1}{m_i}\sum_j\frac{Y_iY_j}{\langle y_i\rangle \langle y_j\rangle}\frac{4k_i k_j}{k_i + k_j}\frac{{\bf r}_{ij}\cdot\nabla \tilde{W}_{ij}}{{\bf r}^2_{ij}}(T_i - T_j).
\end{eqnarray}

\subsection{Viscosity term based on velocity gradient}\label{sec::new_visc}

In this section, we describe our formulation of the viscosity term, which is fundamentally similar to those proposed in various previous works \citep[e.g.,][]{sijacki2006}.

\subsubsection{Viscosity term based on velocity gradient in SSPH}

In this subsection, we describe our novel approach to formulating the viscosity term in SSPH. This approach, which deviates from the pair-wise viscous formulation, involves discretizing the viscous stress tensor $\mathbf{\Pi}$ and incorporating it into the equation of motion, as described in previous works \citep[e.g.,][]{hosono2016b}.

First, we derive the momentum diffusion term in the equation of motion:
\begin{eqnarray}
    \frac{1}{\rho}\nabla\cdot\mathbf{\Pi} = \nabla\cdot\left(\frac{\mathbf{\Pi}}{\rho}\right) + \frac{\mathbf{\Pi}}{\rho^2}\cdot\nabla\rho.
\end{eqnarray}
Consequently, the momentum diffusion term is expressed as
\begin{eqnarray}
    \left(\frac{d\bm{v}_i}{dt}\right)^{\rm visc} &=& \sum_j\frac{m_j}{\rho_j}\frac{\mathbf{\Pi}_j}{\rho_j}\nabla \tilde{W}_{ij} + \frac{\mathbf{\Pi}_i}{\rho_i^2}\sum_j\frac{m_j}{\rho_j}\rho_j\nabla \tilde{W}_{ij} \nonumber \\
    &=& \sum_jm_j \left(\frac{\mathbf{\Pi}_i}{\rho_i^2}+\frac{\mathbf{\Pi}_j}{\rho_j^2}\right)\nabla \tilde{W}_{ij}.
\end{eqnarray}
The internal energy change corresponds to the kinetic energy change, leading to:
\begin{eqnarray}
    m_i\left(\frac{du_i}{dt}\right)^{\rm visc}_{j\to i} + m_j\left(\frac{du_j}{dt}\right)^{\rm visc}_{i\to j}
    &=& -\frac{m_i m_j}{m_i + m_j}\bm{v}_{ij}\cdot\left[\left(\frac{d\bm{v}_i}{dt}\right)^{\rm visc}_{j\to i} - \left(\frac{d\bm{v}_j}{dt}\right)^{\rm visc}_{i\to j}\right] \nonumber \\
    &=& -m_i \bm{v}_{ij}\cdot\left(\frac{d\bm{v}_i}{dt}\right)^{\rm visc}_{j\to i} \nonumber \\
    &=& -m_im_j\bm{v}_{ij} \cdot \left(\frac{\mathbf{\Pi}_i}{\rho_i^2}+\frac{\mathbf{\Pi}_j}{\rho_j^2}\right)\nabla \tilde{W}_{ij}.
\end{eqnarray}
Assuming equal exchanges of internal energy, $m_idu_i/dt|_{j\to i} = m_jdu_j/dt|_{i\to j}$, the viscous heating term becomes
\begin{eqnarray}
    \left(\frac{du_i}{dt}\right)^{\rm visc} = -\sum_j\frac{m_j}{2}\bm{v_{ij}}\cdot\left[\left(\frac{\mathbf{\Pi}_i}{\rho_i^2}+\frac{\mathbf{\Pi}_j}{\rho_j^2}\right)\nabla \tilde{W}_{ij}\right].
\end{eqnarray}
In order to calculate $\mathbf{\Pi}$, we need to derive expressions for $\nabla\cdot\bm{v}$ and $\nabla\otimes\bm{v}$. Using equation (\ref{eq:discretedf_SSPH}), the expression for $\nabla\cdot\bm{v}$ in SPH is
\begin{equation}
    \left(\nabla\cdot\bm{v}\right)_i = \frac{1}{\rho_i}\sum_jm_j(\bm{v}_j - \bm{v}_i)\cdot\nabla W_{ij}(h_i).
\end{equation}
By substituting the operator $\cdot$ with $\otimes$, we obtain the SPH expression for $\nabla\otimes\bm{v}$:
\begin{equation}
    \left(\nabla\otimes\bm{v}\right)_i = \frac{1}{\rho_i}\sum_jm_j(\bm{v}_j - \bm{v}_i)\otimes\nabla W_{ij}(h_i).
\end{equation}

\subsubsection{Viscosity term based on velocity gradient in DISPH}

In this subsection, we outline our approach for formulating the viscous term based on velocity gradient within the DISPH framework. Firstly, we derive the viscous heating term in the energy equation. Given that $\mathbf{\Pi}$ is a symmetrical tensor, it can be expressed as
\begin{eqnarray}
\mathbf{\Pi}:(\nabla\otimes\bm{v}) = \mathbf{\Pi}:(\nabla\otimes\bm{v})^{\mathsf T}.
\end{eqnarray}
Consequently, the viscous heating term is formulated as
\begin{eqnarray}
    \left(\frac{du_i}{dt}\right)^{\rm visc}
    &=& \frac{1}{\rho_i} \mathbf{\Pi}_i:\mathbf{e}_i \nonumber \\ 
    &=& \frac{1}{2\rho_i} \mathbf{\Pi}_i:[(\nabla\otimes\bm{v}_i) + (\nabla\otimes\bm{v}_i)^{\mathsf T}] \nonumber \\
    &=& \frac{1}{\rho_i} \mathbf{\Pi}_i:(\nabla\otimes\bm{v}_i).
\end{eqnarray}
In order to achieve the DISPH formulation for the equation of energy, we need to express $\mathbf{\Pi}:(\nabla\otimes\bm{v})$. This is achieved using
\begin{eqnarray}
    \nabla\cdot(\bm{v}\cdot\mathbf{\Pi}) = \mathbf{\Pi}:(\nabla\otimes\bm{v}) + \bm{v}\cdot(\nabla\cdot\mathbf{\Pi}).
\end{eqnarray}
The formulation of $\mathbf{\Pi}:(\nabla\otimes\bm{v})$ is then given by
\begin{eqnarray}
    \mathbf{\Pi}_{i}\cdot(\nabla\otimes\bm{v})_{i}
    &=& \nabla\cdot(\bm{v}_i\cdot\mathbf{\Pi}_i) - \bm{v}_i\cdot(\nabla\cdot\mathbf{\Pi}_i) \nonumber \\
    &=& \sum_j \bm{v}_j\cdot\mathbf{\Pi}_j\frac{Y_j}{\langle y_j\rangle}\nabla W(\bm{r}_{ij};h_i) - \bm{v}_i\cdot\sum_j\mathbf{\Pi}_j\frac{Y_j}{\langle y_j\rangle}\nabla W(\bm{r}_{ij};h_i) \nonumber \\
    &=& -\sum_j\frac{Y_j}{\langle y_j\rangle}\bm{v}_{ij}\cdot[\mathbf{\Pi}_j \cdot \nabla W(\bm{r}_{ij};h_i)]. \label{dummy}
\end{eqnarray}
From equation (\ref{dummy}), we obtain the novel viscous heating term:
\begin{eqnarray}
    \left(\frac{du_i}{dt}\right)^{\rm visc} = - \frac{1}{m_i}\sum_j \frac{Y_i Y_j}{\langle y_i\rangle\langle y_j\rangle} \bm{v}_{ij} \cdot [\mathbf{\Pi}_j \cdot \nabla W(\bm{r}_{ij};h_i)]. \label{vischeat_VG}
\end{eqnarray}
The internal energy change is equivalent to the kinetic energy change, as shown:
\begin{eqnarray}
    m_i\left(\frac{du_i}{dt}\right)^{\rm visc}_{j\to i} + m_j\left(\frac{du_j}{dt}\right)^{\rm visc}_{i\to j}
    &=& -\frac{m_i m_j}{m_i + m_j}\bm{v}_{ij}\cdot\left[\left(\frac{d\bm{v}_i}{dt}\right)^{\rm visc}_{j\to i} - \left(\frac{d\bm{v}_j}{dt}\right)^{\rm visc}_{i\to j}\right], \nonumber \\
    &=& -m_i \bm{v}_{ij}\cdot\left(\frac{d\bm{v}_i}{dt}\right)^{\rm visc}_{j\to i}.
    \label{dummy2}
\end{eqnarray}
From equation (\ref{vischeat_VG}), we derive
\begin{eqnarray}
    m_i\left(\frac{du_i}{dt}\right)^{\rm visc}_{j\to i} &+& m_j\left(\frac{du_j}{dt}\right)^{\rm visc}_{i\to j} \nonumber \\
    &=& - \frac{Y_i Y_j}{\langle y_i\rangle\langle y_j\rangle} \bm{v}_{ij} \cdot [\mathbf{\Pi}_j \cdot \nabla W(\bm{r}_{ij};h_i)] - \frac{Y_i Y_j}{\langle y_i\rangle\langle y_j\rangle} \bm{v}_{ji} \cdot [\mathbf{\Pi}_i \cdot \nabla W(\bm{r}_{ji};h_j)], \nonumber \\
    &=& - \frac{Y_i Y_j}{\langle y_i\rangle\langle y_j\rangle} \bm{v}_{ij} \cdot [\mathbf{\Pi}_j \cdot \nabla W(\bm{r}_{ij};h_i) + \mathbf{\Pi}_i \cdot \nabla W(\bm{r}_{ij};h_j)]. \label{dummy3}
\end{eqnarray}
By applying equations (\ref{dummy2}) and (\ref{dummy3}), we finally formulate the momentum diffusion term:
\begin{eqnarray}
\left(\frac{d\bm{v}_i}{dt}\right)^{\rm visc} = \sum_j \frac{1}{m_i} \frac{Y_i Y_j}{\langle y_i\rangle\langle y_j\rangle} [\mathbf{\Pi}_j \cdot \nabla W(\bm{r}_{ij};h_i) + \mathbf{\Pi}_i \cdot \nabla W(\bm{r}_{ij};h_j)].
\end{eqnarray}

\subsection{Radiative cooling} \label{sec:radiative}

In our model, we assume that radiation emitted from the satellite's surface follows blackbody radiation principles. The radiative cooling term in the energy equation is represented as
\begin{eqnarray}
    \left(\frac{du_i}{dt}\right)^{\rm rad} = - \omega_i\frac{\sigma T_{\rm surface}^4 S_{\rm surface}}{m_i} \hspace{20pt}\mbox{(\rm for surface particles)},
\end{eqnarray}
where $\omega_i$ is the weighting factor, $T_{\rm surface}$ is the surface temperature of the icy moon, and $S_{\rm surface}$ is the surface area. We assume that the tidal deformation-induced bulge is relatively small compared to the satellite's radius, leading to
\begin{equation}
    S_{\rm surface} = 4 \pi R_{\rm sat}^2,
\end{equation}
where $R_{sat}$ is the radius of the satellite.

\begin{figure}
    \begin{center}
        \includegraphics[width=100mm]{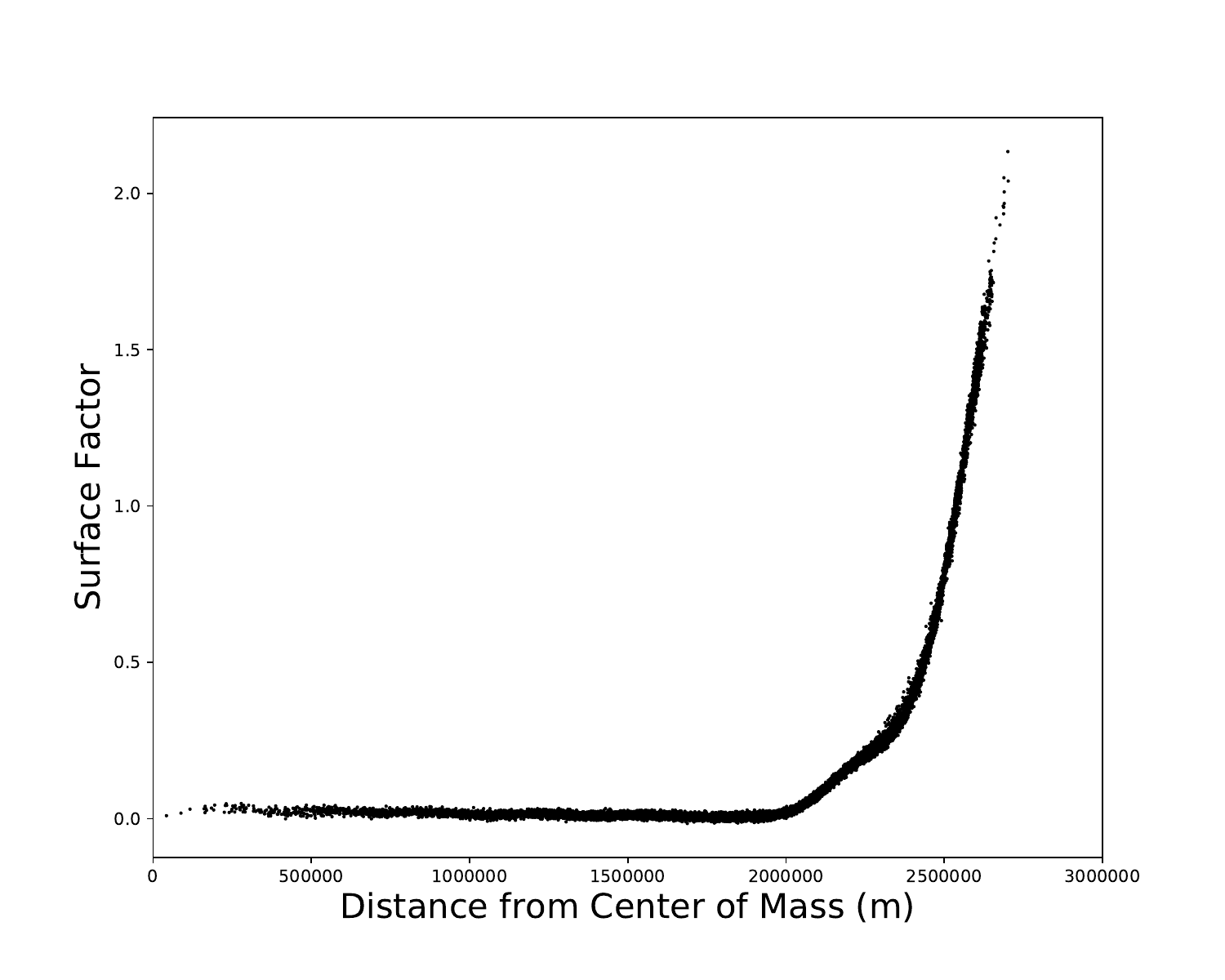}
    \end{center}
    \caption{The radial profile of Surface factor}
    \label{fig:surfacefactor}
\end{figure}

\begin{figure}
    \begin{center}
        \includegraphics[width=100mm]{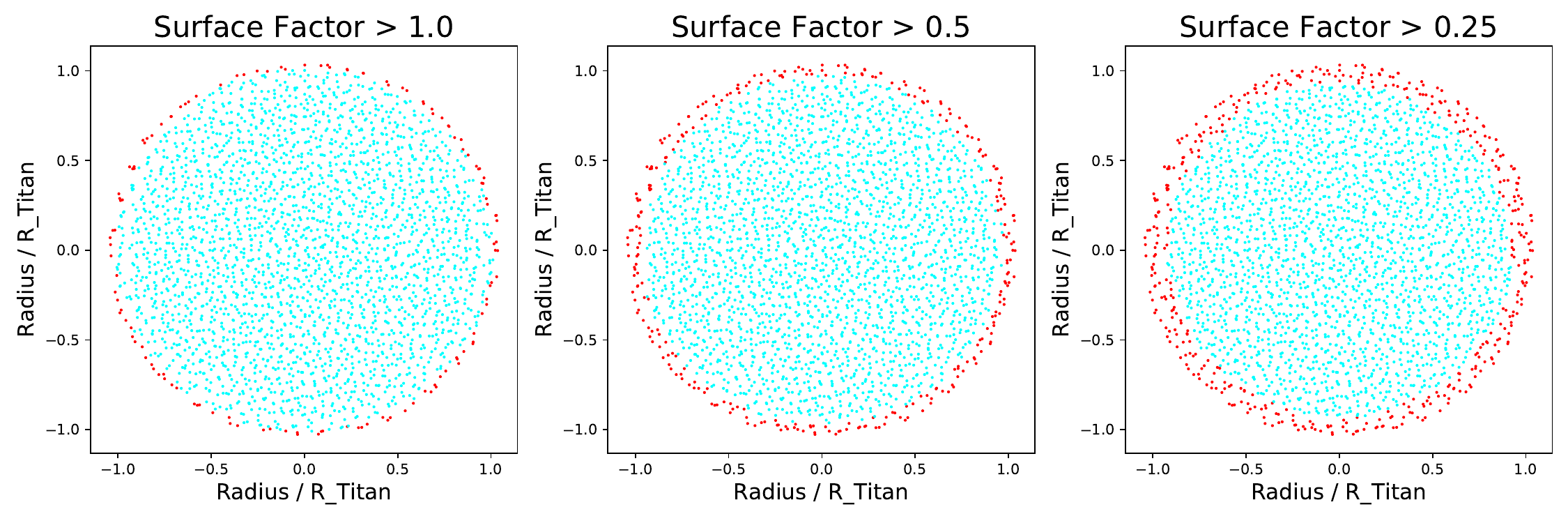}
    \end{center}
    \caption{Surface particles are shown in red for the threshold $>1.0$ (left), $>0.5$ (middle) and $>0.25$ (right.)}
    \label{fig:surface}
\end{figure}

The ``surface factor'' ($F_{\rm surface}$) is defined as the normalized distance between the coordinates of an SPH particle and the center of gravity of surrounding particles (Figure \ref{fig:surfacefactor}):
\begin{eqnarray}
    F_{\rm surface} = \left|\bm{r}_i - \frac{\Sigma_j \bm{r}_j W_{ij}}{\Sigma_j W_{ij}}\right|\times\left(\frac{3 m_i}{4\pi\rho_i}\right)^{-\frac{1}{3}}.
\end{eqnarray}
Particles with a surface factor exceeding a specified threshold are designated as ``surface particles'' (Figure \ref{fig:surface}). We have set this threshold at 0.5. The surface temperature, $T_{\rm surface}$, is calculated based on the surface factor and the temperatures of surface particles through weighted averaging. Radiative cooling is then distributed from these surface particles, weighted by the square of their surface factor.

\subsection{Timestep}\label{sec:timestep}

In our simulations, the timesteps constrained by momentum and thermal diffusion are defined as follows
\begin{eqnarray}
    \Delta t_i^{\rm M} &=& C^{\rm M} \frac{h_i^2}{\nu_i}, \\
    \Delta t_i^{\rm T} &=& C^{\rm T} \frac{h_i^2}{\kappa_i},
\end{eqnarray}
where $\nu$ represents the kinematic viscosity, and $\kappa$ is the thermal diffusivity. We have chosen coefficients $C^{\rm M}$ and $C^{\rm T}$ to be 0.1. The timestep constrained by the CFL condition is as per equation (\ref{timestepCFL}). The minimum of these timestep constraints is applied to our simulation:
\begin{eqnarray}
    \Delta t_i = \min_i\left(\Delta t_i^{\rm CFL}, \Delta t_i^{\rm M}, \Delta t_i^{\rm T}\right).
\end{eqnarray}

However, $\Delta t^{\rm M}$ tends to be very small due to the high viscosity of the ice shell, which also varies strongly with temperature as shown in equation (\ref{viscmodel}). This variability poses a challenge for simulations. To address this, we employ the Variable Inertia Method (VIM) as proposed by \cite{takeyama2017}. VIM transforms the basic equations (equations (\ref{eqmotion}) and (\ref{eqenergy})) using parameters $\xi$, $\chi$, and $\phi$:
\begin{eqnarray}
    &&\frac{d\bm{v}}{dt} = \frac{1}{\xi}\left(\frac{\phi}{\rho}\nabla p + \frac{1}{\chi\rho}\nabla \cdot \mathbf{\Pi} + \bm{g}\right), \\
    &&\frac{du}{dt} = - \frac{p}{\rho}\nabla\cdot\bm{v} + \frac{\phi}{\chi\rho}\mathbf{\Pi}:\mathbf{e} +\nabla\cdot(\chi k\nabla T) - \dot{u}^{\mathsf rad}.
\end{eqnarray}
By adjusting $\xi$, $\chi$, and $\phi$, we can modify the Reynolds number, Mach number, and Prandtl number, respectively, while keeping the Rayleigh number unchanged. Under VIM, the sound speed, kinematic viscosity, and thermal diffusivity are altered to
\begin{eqnarray}
    &&c^{\prime} = \frac{1}{\sqrt{\xi}}c, \\
    &&\nu^{\prime} = \frac{\phi}{\xi \chi}\nu, \\
    &&\kappa^{\prime} = \chi \kappa.
\end{eqnarray}
As a result, the timestep is changed to
\begin{eqnarray}
    \Delta t_i = \min_i\left(\sqrt{\xi}\Delta t_i^{\rm CFL}, \frac{\xi\chi}{\phi}\Delta t_i^{\rm M}, \frac{1}{\chi}\Delta t_i^{\rm T}\right).
\end{eqnarray}
According to the \cite{takeyama2017}, we set $\phi = 1$, $\xi = 1.0 \times 10^6$, and $\chi = 1.0 \times 10^3$.

\subsection{Discretized governing equations}

Building upon the previously discussed formulations, the discretized equations of motion and energy for icy moons, incorporating our novel viscosity formulation in DISPH, are expressed as
\begin{eqnarray}
    \frac{d\bm{v}_i}{dt} &=& - \frac{1}{\xi}\sum_j \frac{Y_i Y_j}{m_i}\left[\frac{\langle y_i\rangle^{1/\alpha-2}}{\Omega_i}\nabla W_{ij}(h_i) + \frac{\langle y_j\rangle^{1/\alpha-2}}{\Omega_j}\nabla W_{ij}(h_j)\right] \nonumber \\
    &&\hspace{5pt}+ \sum_j \frac{1}{m_i} \frac{Y_i Y_j}{\langle y_i\rangle\langle y_j\rangle} [\mathbf{\Pi}^{\prime}_j \cdot \nabla W(\bm{r}_{ij};h_i) + \mathbf{\Pi}^{\prime}_i \cdot \nabla W(\bm{r}_{ij};h_j)], \\
    \frac{du_i}{dt} &=&\sum_j \frac{Y_i Y_j}{m_i}\frac{\langle y_i\rangle^{1/\alpha-2}}{\Omega_i}\bm{v}_{ij}\cdot\nabla W_{ij}(h_i) - \xi\frac{1}{m_i}\sum_j \frac{Y_i Y_j}{\langle y_i\rangle\langle y_j\rangle} \bm{v}_{ij} \cdot [\mathbf{\Pi}^{\prime}_j \cdot \nabla W(\bm{r}_{ij};h_i)] \nonumber \\
    && \hspace{10pt}+ \frac{1}{m_i}\sum_j\frac{Y_iY_j}{\langle y_i\rangle \langle y_j\rangle}\frac{4k^{\prime}_i k^{\prime}_j}{k^{\prime}_i + k^{\prime}_j}\frac{{\bf r}_{ij}\cdot\nabla \tilde{W}_{ij}}{{\bf r}^2_{ij}}(T_i - T_j) - \omega_i\frac{\sigma T_{\rm surface}^4 S_{\rm surface}}{m_i}, \label{eq:energy}
\end{eqnarray}
where $\mathbf{\Pi}^{\prime} = \mathbf{\Pi} / (\xi\chi)$ and $k^{\prime} = \chi k$. These equations integrate the modified viscosity and thermal conductivity terms, alongside the radiative cooling model, providing a comprehensive framework for simulating the dynamics and thermal evolution of icy moons.

\section{Numerical test}\label{sec:results}
In order to evaluate our novel formulation against the conventional approach, we simulated a homogeneous icy object approximating the size of Titan, setting its rotation to match synchronous rotation at a distance 1.1 times the Roche radius from Saturn.
This object has a radius of about 2500 km and a mass of roughly $7\times10^{22}$ kg. For the simulation, we employed $\sim 3\times10^4$ SPH particles. In these calculations, the icy moon's self-gravity was accounted for as the gravitational force in equation (2).
We used Tillotson's EOS, as phase transitions do not influence these results. We have two options for the viscous term and the formulation of SPH method.

The validation of our new viscosity formulation in simulations to test its ability to capture physically realistic fluid behaviour is displayed in \ref{sec:HydrodynamicTests}. We also show a comparison of computational runtimes between the traditional DISPH formulation and our newly proposed scheme in the hydrodynamic tests in \ref{sec:computationtime}

\subsection{Modification of viscous formulation}
\label{sec:ViscModification}

First, we present a comparison between the conventional and our novel formulation of viscosity in SSPH.

\subsubsection{Conventional formulation vs. our formulation}

\begin{figure}
    \begin{center}
        \includegraphics[width=100mm]{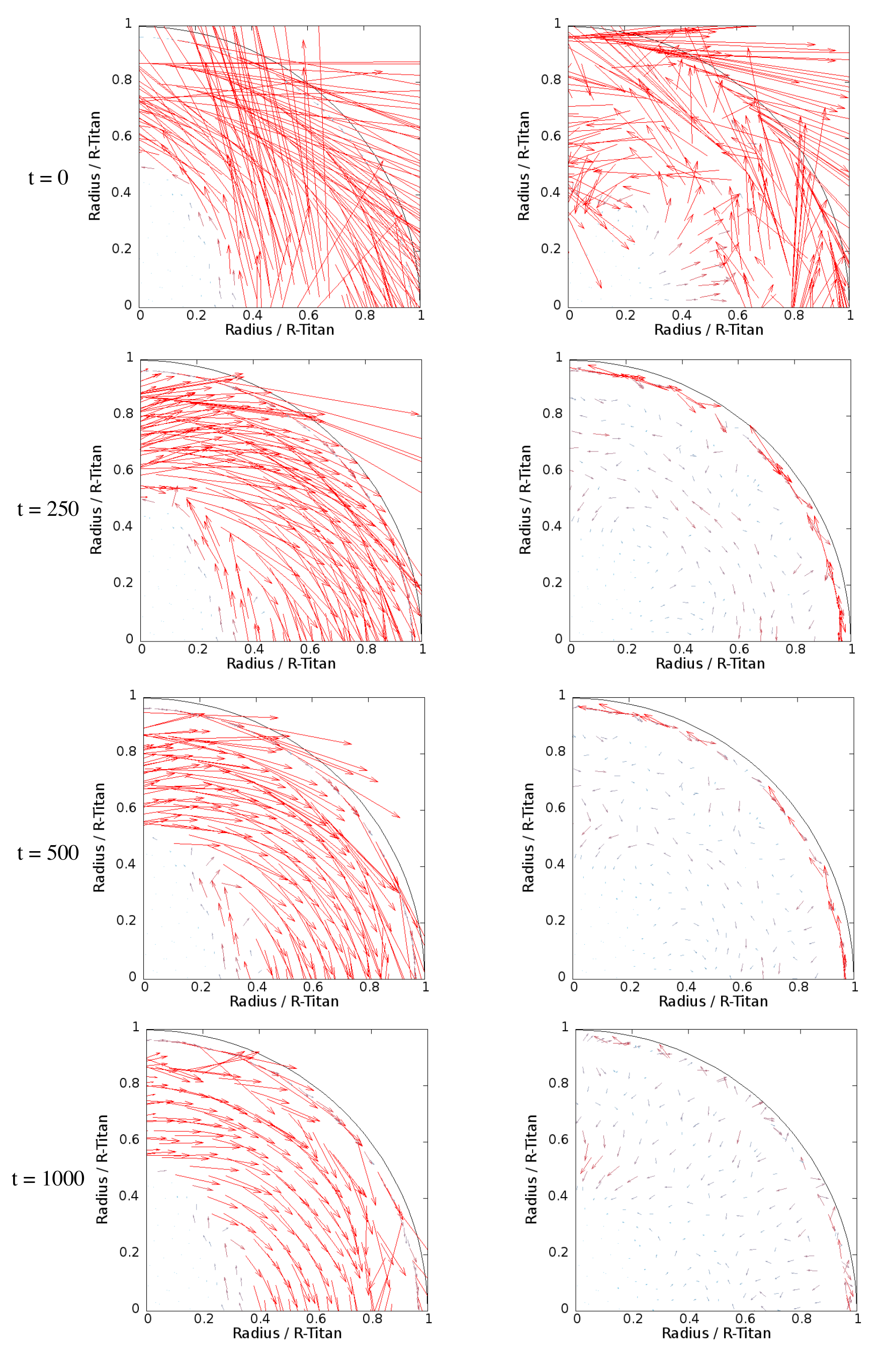}
    \end{center}
    \caption{The acceleration vector caused by the viscous force for  formulation of viscosity based on the relative velocity (left) and the velocity gradient (right). In this calculation, we use the Tillotson EOS.}
    \label{fig:vector_acc}
\end{figure}

We encountered an issue when simulating rigid body rotation using the conventional viscous term in SPH. With this formulation, an unintended viscous force inhibits rotation, as illustrated in the left panels of Figure \ref{fig:vector_acc}.

This occurs because the conventional viscosity term is predicated on relative velocity (equations (\ref{eq:viscforce_RV}) and (\ref{eq:vischeat_RV})). In a rigid body rotation scenario, viscous forces cause particles to be accelerated by outer particles and decelerated by inner particles. However, particles on the surface only experience deceleration due to the absence of outer particles, leading to an overall slowing down of the entire satellite. This phenomenon results in the deceleration of surface particles. Consequently, the force acting on the particles just beneath the surface, in the direction of deceleration, becomes stronger than the force driving acceleration. This effect propagates through the satellite, leading to an overall slowing down of its rotation.

Conversely, the right panels in Figure \ref{fig:vector_acc} show acceleration vectors using our new viscous formulation. This approach, based on the velocity gradient, effectively suppresses the unwanted forces, demonstrating the advantage of our novel formulation.
The discussion about conservation features is described in \ref{sec:conservations}.

\subsection{Modification of heat distribution} \label{sec:HeatModification}

\begin{figure}
    \begin{center}
        \includegraphics[width=100mm]{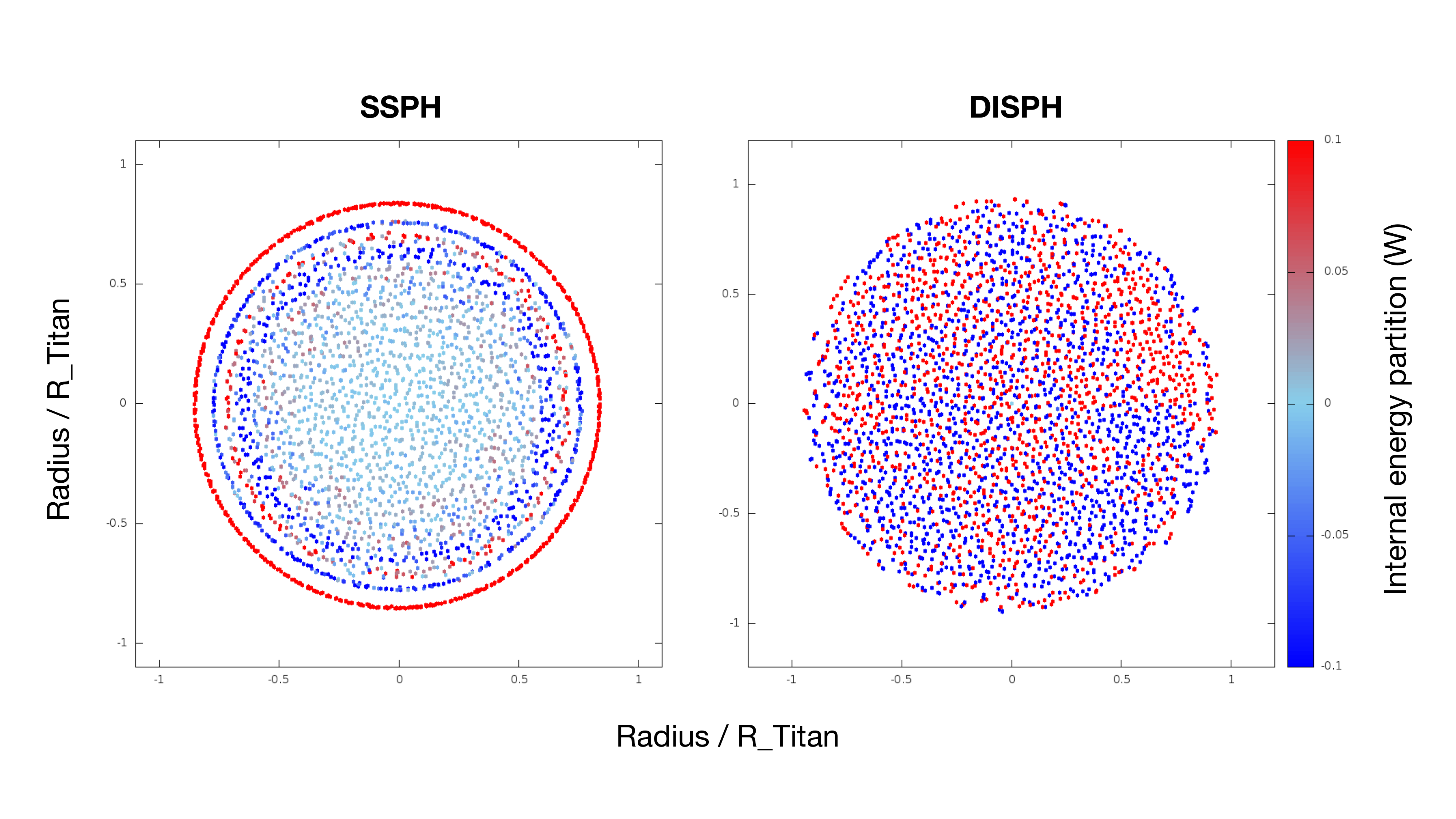}
    \end{center}
    \caption{The internal energy change rate depend on the viscosity for the formulation based on velocity gradient in SSPH (left) and DISPH (right). This temperature distribution is not time-varying.}
    \label{fig:internalenergy}
\end{figure}

Next, we present a comparative analysis of SSPH and DISPH utilizing our novel viscosity formulation. As depicted in the left panel of Figure \ref{fig:internalenergy}, the internal energy change rate, influenced by viscosity within the velocity gradient-based formulation, reveals an artificial layered structure in the energy distribution. This characteristic structure is specific to SSPH, which operates under the assumption of a differentiable density throughout the medium.

To address this limitation, we incorporated DISPH, a method developed by \cite{saitoh2013}, \cite{hosono2013a}, and \cite{hopkins2013}.
Unlike SSPH, DISPH ensures differentiability in pressure rather than density.
This adjustment significantly improves behavior at free boundaries and contact discontinuity surfaces.

The right panel of Figure \ref{fig:internalenergy} illustrates the internal energy change rate dependent on viscosity in the velocity gradient-based formulation in DISPH. While SSPH exhibits an artificial layered structure in energy distribution, DISPH effectively resolves this issue.

\section{Conclusion} \label{sec:conclusion}

Understanding the structure and evolution of internal oceans, where the presence of liquid water is suggested, is crucial for astrobiological studies. To this end, we aimed to simulate these environments through 3-dimensional numerical fluid calculations using the SPH method. We developed a comprehensive SPH method code that includes viscosity, heat conduction, radiative cooling, and phase transition mechanisms.

During our study, we encountered two primary issues when using the SPH method to simulate a rigid body undergoing rotational motion with viscosity:
(1) The conventional formulation of the viscosity term unphysically halts the rotation of the rigid body.
(2) The standard SPH creates artificial internal energy partitioning, leading to the formation of artificial layers of SPH particles. To overcome these challenges, we implemented two key modifications:
(a) We reformulated the viscosity term based on the velocity gradient.
(b) We enhanced the SPH scheme by introducing DISPH. These modifications successfully resolved the issues associated with simulating spinning viscous fluids.

Furthermore, we introduced a method for radiative cooling by identifying 'surface particles' of a planet. This was achieved by using particle coordinates and calculating the distance from the center of gravity based on surrounding particles as an index.

As future works, we are planning to perform two patterns of calculations:
(1) Applying our model to hypothetical systems for a parameter study. We plan to simulate various scenarios by varying semi-major axis, eccentricity, obliquity, mass of the icy moon, etc., to analyze their effects on the internal structure.
(2) Applying our model to actual icy moons in the solar system, such as Europa and Enceladus, to uncover and understand their internal structures.

\section*{Acknowledgements}
=We thank the anonymouns reviewer's constructive comments, which led us to greatly improve this paper. This work was supported by JSPS KAKENHI Grant Number 21H04512 and Grant-in-Aid for JSPS Research Fellow Number 22J22428.

\appendix

\section{Hydrodynamic Tests}\label{sec:HydrodynamicTests}

In this section, we validate of our new viscosity formulation in simulations to test its ability to capture physically realistic fluid behaviour.
We perform Hydrostatic Equilibrium tests, Kelvin-Helmholtz Instability tests and Rayleigh-Taylor Instability tests using SSPH with traditional formulation of viscosity, DISPH with traditional formulation of viscosity and DISPH with our new formulation of viscosity.
We confirm that hydrodynamic tests that cannot be correctly computed with SSPH but DISPH with traditional viscosity formulation can gives better solution can also be solved correctly using our new viscosity formulation.
All tests are calculated in 2D simulation.
In the following calculations, we use an ideal gas EoS with $\gamma = 5.0 /3.0$ and set the viscosity $\mu = 2.5\times 10^{-5}$ and the thermal conductivity $k = 0.1682$.

\subsection{Hydrostatic Equilibrium Tests}
We follow the evolution of two fluids of different densities at the same pressure.
The initial conditions are
\begin{eqnarray}
    &&\rho = \begin{cases}
        4 & 0.25 \geq x \geq 0.75 \text{and} \geq x \geq 0.75 \\
        1 & \text{otherwise}, \\
    \end{cases} \\
    &&p = 2.5.
\end{eqnarray}
Particles in the high-density region have four times the mass of particles in the low-density region.
The box has dimensions of 1, 1 in the $x$ and $y$ directions.

\begin{figure}[h]
    \begin{center}
        \includegraphics[width=120mm]{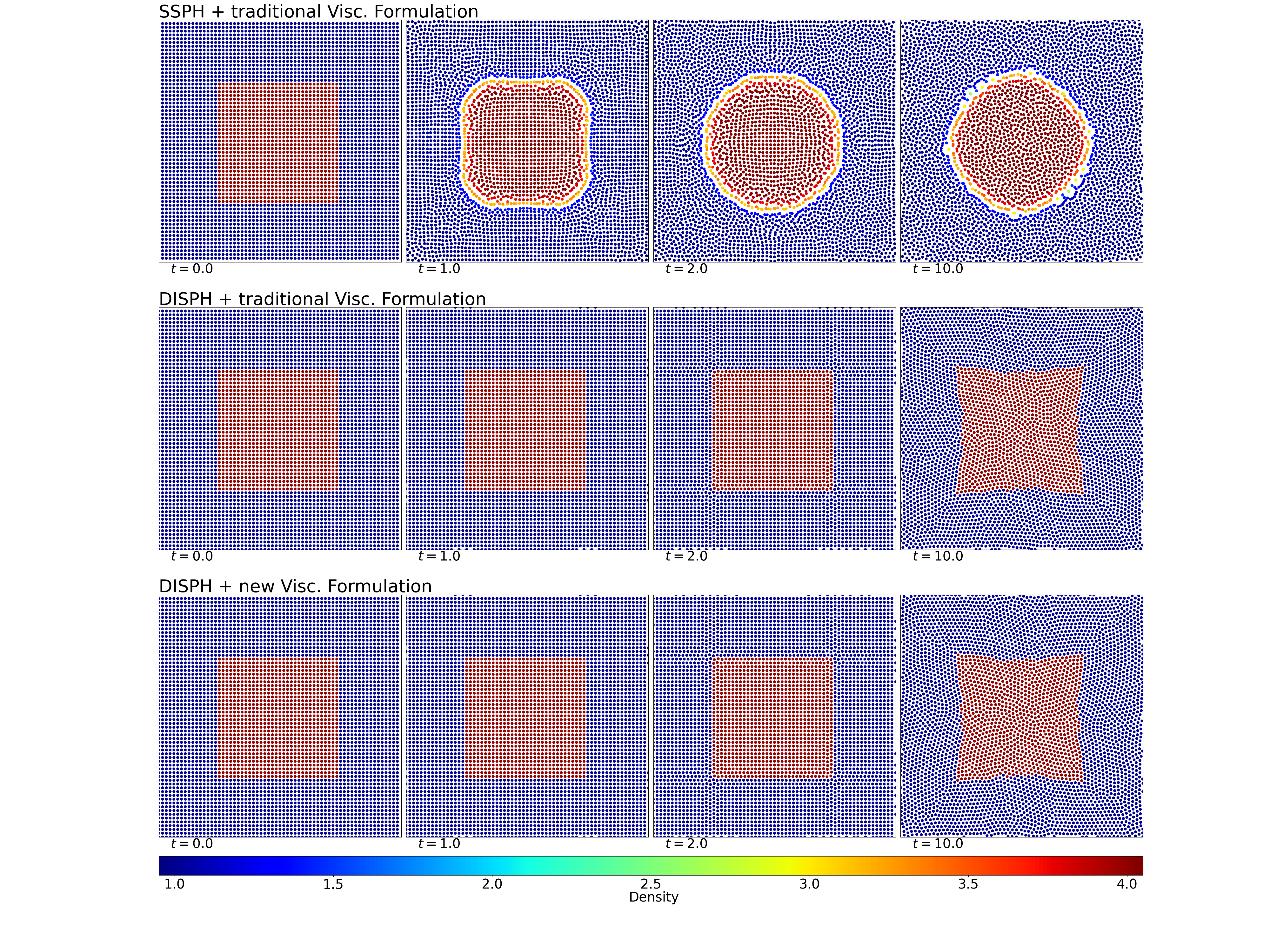}
    \end{center}
    \caption{Density maps of the two-dimensional Hydrostatic equilibirium tests at t = $0.0, 1.0, 2.0$, and $10.0$. 
    The upper panels show the results of the SSPH + traditional viscosity term, the middle panels show those of DISPH + traditional viscosity term and the bottom panels show those of DISPH + our new viscosity term.}
    \label{fig:HS}
\end{figure}

Figure \ref{fig:HS} shows the time evolution up to $t = 10.0$.
As already pointed out from \citet{saitoh2013}, there is a clear difference between the result of the SSPH and that of DISPH.
In SSPH, high-density areas that are initially square in shape quickly become round.
On the other hand, the overall square shape remains until the end of the simulation in DISPH.

\subsection{Kelvin-Helmholtz Instability Tests}\label{sec:KHItest}
Following \citet{tricco2019kelvin}, the initial conditions are given by
\begin{eqnarray}
    &&\rho = 1 + \frac{1}{2}\left[\tanh\left(\frac{y-0.5}{a}\right) - \tanh\left(\frac{y-1.5}{a}\right) \right], \\
    &&v_x = v_0\left[\tanh\left(\frac{y-0.5}{a}\right) - \tanh\left(\frac{y-1.5}{a}\right) - 1.0\right], \\
    &&v_y = A\sin (2\pi x)\left[\exp\left(-\frac{(y-0.5)^2}{\sigma^2}\right) + \exp\left(-\frac{(y-1.5)^2}{\sigma^2}\right)\right], \\
    &&p = 10.0,
\end{eqnarray}
where $a = 0.05$ and $\sigma = 0.2$.
We take $v_0 = 1.0$ and $A = 0.01$.
The box has dimensions of 2, 2 in the $x$ and $y$ directions.
The periodic boundary condition is imposed on the $x$ and $y$ directions.
The total number of particles is $262144$ and the particle mass is set to $2.3 \times 10^{-5}$.
We use the stretch mapping technique \citep{price2018phantom} to achieve the correct density profile.

\begin{figure}[h]
    \begin{center}
        \includegraphics[width=120mm]{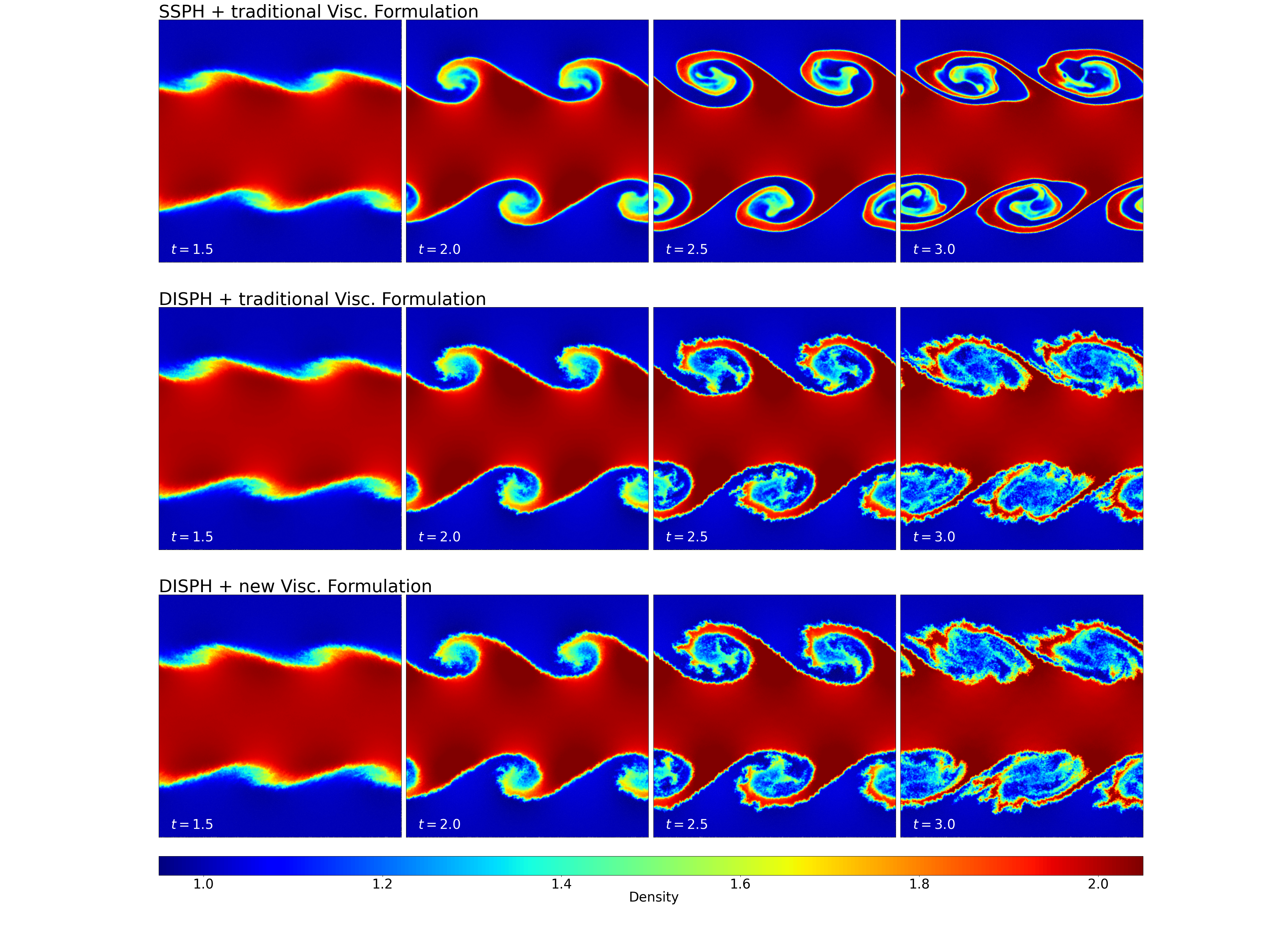}
    \end{center}
    \caption{Density maps of the two-dimensional Kelvin-Helmholtz instability tests at t = $1.0, 1.5, 2.0$, and $2.5$. 
    The upper panels show the results of the SSPH + traditional viscosity term, the middle panels show those of DISPH + traditional viscosity term and the bottom panels show those of DISPH + our new viscosity term.}
    \label{fig:KHI}
\end{figure}

In Figure \ref{fig:KHI}, we show snapshots from KHI simulations simulated using SSPH with traditional viscosity term, DISPH with traditional viscosity term and DISPH with our new viscosity term.
The SSPH simulation shows that effects like surface tension work to suppress the growth of instabilities and prevent particle growing of roll-like structures
On the other hand, DISPH with traditional viscosity term and DISPH with new viscosity term show very good results which are comparable to \citet{mcnally2012}.

\begin{figure}[h]
    \begin{center}
        \includegraphics[width=120mm]{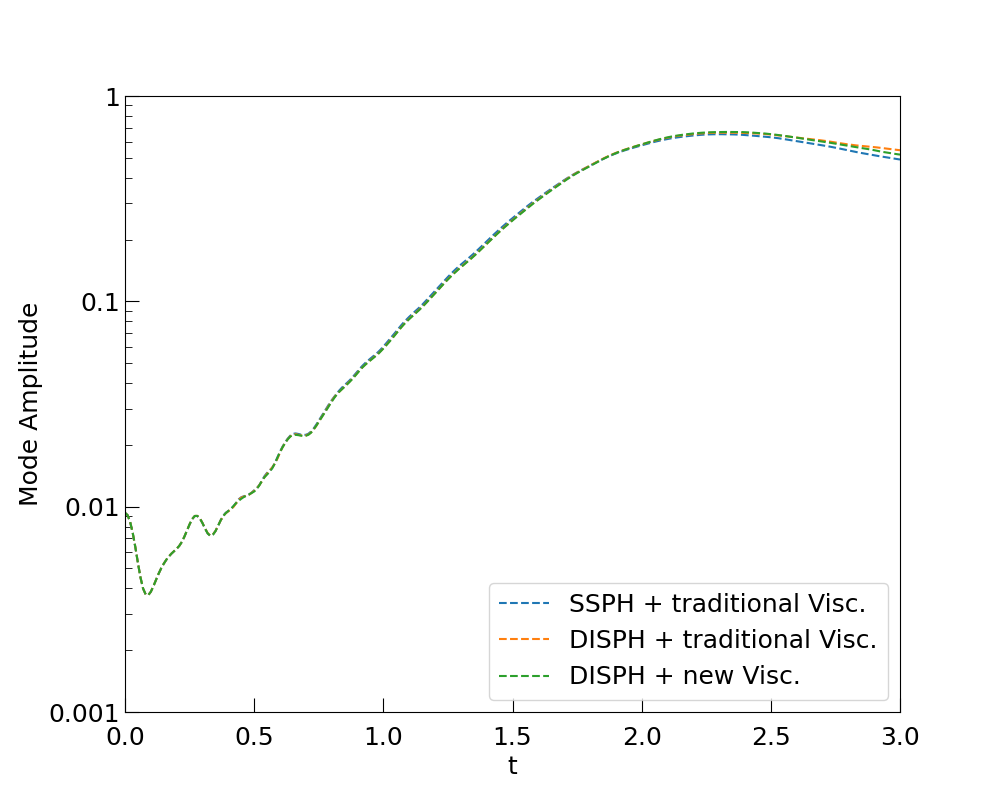}
    \end{center}
    \caption{The time evolution of Kelvin–Helmholtz instability mode amplitude. Mode growth in simulations using both SSPH with traditional viscosity term (blue, dashed), DISPH with traditional viscosity term (orenge, dashed) and DISPH with new viscosity term (green, dashed) are shown.}
    \label{fig:KHI_amp}
\end{figure}

Figure \ref{fig:KHI_amp} shows the mode amplitude ($M$) growth of these KHIs.
The mode amplitude is calculated based on a discrete convolution, as described by \citet{mcnally2012}.
The mode amplitude is given by
\begin{eqnarray}
    &&d_i = \begin{cases}
        h_i^2 \exp(- |y_i - 0.5| / \sigma^2) & y < 1 \\
        h_i^2 \exp(- |(2 - y_i) - 0.5| / \sigma^2) & y \geq 1,
    \end{cases}\\
    &&s_i = v_{y,i} \sin(2\pi x_i),\\
    &&c_i = v_{y,i} \cos(2\pi x_i),\\
    &&M = 2\left[\left(\frac{\sum^N_{i=1}s_i d_i}{\sum^N_{i=1} d_i}\right)^2 + \left(\frac{\sum^N_{i=1}c_i d_i}{\sum^N_{i=1} d_i}\right)^2\right]^{1/2},
\end{eqnarray}
where $i$ is index of each particle and $N$ is the number of particles.
The mode amplitude in all simulations are in good agreement with Figure 2 in \citet{tricco2019kelvin}.

\subsection{Rayleigh-Taylor Instability Tests}\label{sec:RTItest}

\begin{figure}
    \begin{center}
        \includegraphics[width=120mm]{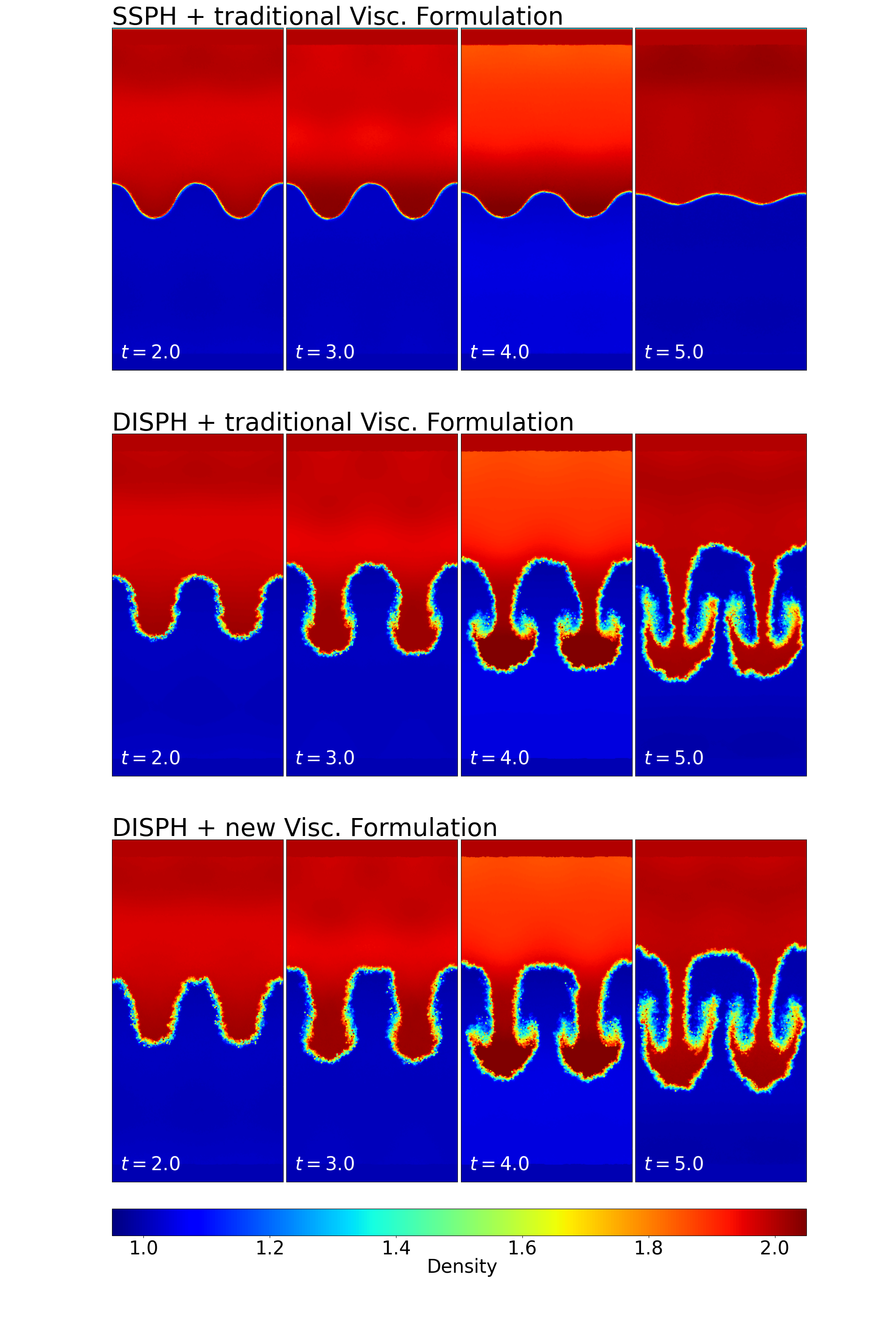}
    \end{center}
    \caption{Density maps of the two-dimensional Rayleigh–Taylor instability tests at t = $2.0, 3.0, 4.0$, and $5.0$. 
    The upper panels show the results of the SSPH + traditional viscosity term, the middle panels show those of DISPH + traditional viscosity term and the bottom panels show those of DISPH + our new viscosity term.}
    \label{fig:RTI}
\end{figure}

Next, we consider the Rayleigh–Taylor instability (RTI).
The box has dimensions of 0.5, 1 in the $x$ and $y$ directions.
The low density region with density $\rho_1 = 1$ occupies the bottom half of the domain, while the bottom half of the domain is occupied by high density region with density $\rho_2 = 2$.
We added the velocity perturbation to particles as follows:
\begin{eqnarray}
    v_y(x,y) = \begin{cases}
        \delta_{vy}[1 + \cos (8\pi (x+0.25))][1+\cos(5\pi (y-0.5))] &\text{for}\hspace{5pt}0.3 < y < 0.7 \\
        0 & \text{otherwise,}
    \end{cases}
\end{eqnarray}
where $\delta_{vy} = 0.025$.
We set initial internal energies to satisfy hydrostatic equilibrium using an ideal gas EoS with $\gamma = 5.0 /3.0$.
The constant acceleration of gravity ($g$) is set to $-0.1$ and a pressure at the interface ($P_0$) is set to $\rho_2/\gamma$.
The total number of particles is $131072$ and the particle mass is set to $5.7 \times 10^{-6}$.
Particles with $y < 0.05$ and $y > 0.95$ are fixed and the periodic boundary condition is imposed on the $x$ direction.
We use the stretch mapping technique \citep{price2018phantom} as we did for the KHI calculations.

In Figure \ref{fig:RTI}, we show snapshots from RTI simulations.
The RTI develops in calculations with the DISPH with traditional viscosity term and DISPH eith traditional viscosity term, while in the SSPH with traditional viscosity simulation surface tension-like force strongly suppress the growth of instabilities.

\section{Comparison of computation time} \label{sec:computationtime}

\begin{table}
    \begin{center}
        \label{tab:comptime}
        \begin{tabular}{lcc}
        \hline
        & Kelvin-Helmholtz Instability & Rayleigh-Taylor Instability \\
        & (END TIME = 3) & (END TIME = 5)\\
        \hline
        traditional visc. formulation & 6248 sec & 18764 sec  \\
        new visc. formulation & 6482 sec & 19232 sec  \\\hline
        \end{tabular}
        \caption{The comparison the computation time between traditional formlation and our new formulation of viscosity term in DISPH.}
    \end{center}
\end{table}

In this section, we compare the computation time between traditional formulation and our new formulation of viscosity term.
Table B.2 shows the two comparisons of the computation time of Kelvin-Helmholtz instability test (\ref{sec:KHItest}) and Rayleigh-Taylor instability test (\ref{sec:RTItest}) in DISPH.
These calculations were performed using 12 threads of openMP parallel.
The new viscosity formulation requires the use of tensors, unlike the conventional formulation, which slightly increases the calculation run time, but the increase is only a few percent, and the simulation can be performed at almost the same computational cost as the conventional formulation.

\section{Conservation features}\label{sec:conservations}

\subsection{Galilean invariance}
It can be seen that the new viscosity formulation introduces velocity into the energy equation (Equation (\ref{eq:energy})) in the form of relative velocities, which guarantees Galilean invariance.

\subsection{Conservation of linear/angular momemtum}\label{sec:cons_momemtum}
The viscous force from j-particle to i-particle based on the relative velocity $(\bm{F}_{ij}^{\rm viscRV})$ and the velocity gradient $(\bm{F}_{ij}^{\rm viscVG})$ are written as follows:
\begin{eqnarray}
    &&\bm{F}_{ij}^{\rm viscRV} = \frac{m_im_j}{\rho_i \rho_j}\frac{4\mu_i \mu_j}{\mu_i + \mu_j}\frac{\bm{r_{ij}}\cdot\nabla \tilde{W}_{ij}}{\bm{r_{ij}}^2}\bm{v}_{ij}, \\
    &&\bm{F}_{ij}^{\rm viscVG} = m_im_j\left(\frac{{\mathbf \Pi}_i}{\rho_i^2} + \frac{{\mathbf \Pi}_j}{\rho_j^2}\right)\nabla \tilde{W}_{ij}.
\end{eqnarray}
Since $\bm{F}_{ij}^{\rm viscVG}$ is antisymmetric, momentum conservation is guaranteed for viscous forces with the new formulation.

\begin{figure}
    \begin{center}
        \includegraphics[width=80mm]{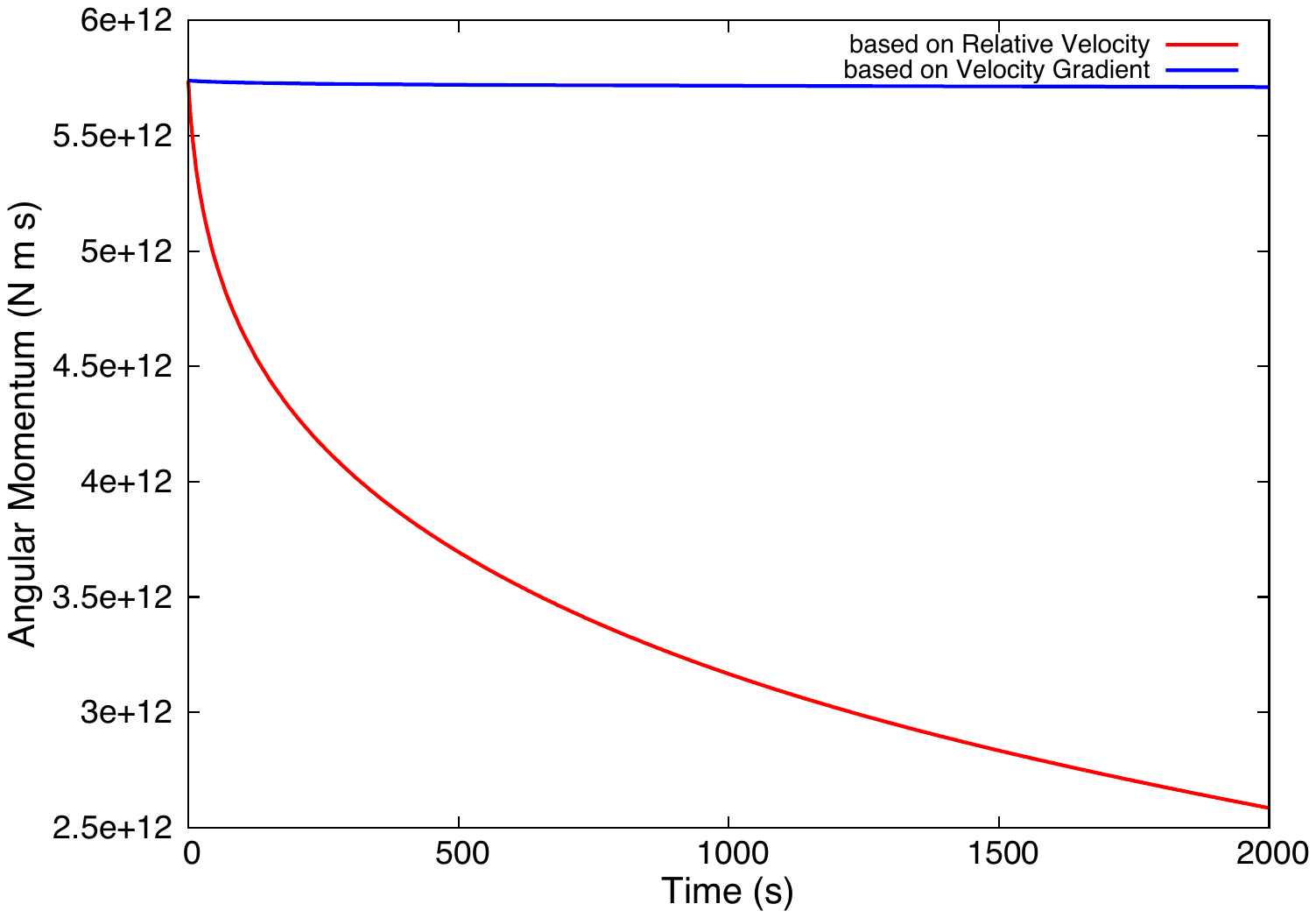}
    \end{center}
    \caption{The time evolution of angular momentum in the numerical test (section \ref{sec:results}) for the formulation based on the relative velocity (red) and the velocity gradient (blue).}
    \label{fig:AngMom}
\end{figure}

The necessary condition for conservation of angular momentum is written as
\begin{eqnarray}
    \bm{r}_i \times \bm{F}_{ij} + \bm{r}_j \times \bm{F}_{ji} = 0 \,\,\,\,\,\,\mbox{for each couple of ($i ,j$)},
\end{eqnarray}
where $\bm{F}$ is the internal force.
$\bm{F}_{ij}^{\rm viscVG}$ and $\nabla \tilde{W}_{ij}$ are antisymmetric, therefore when the force and the relative position vector are collinear, the condition is satisfied.
In the case of $\bm{F}_{ij}^{\rm viscRV}$, the force is parallel to the relative velocity vector, so that conservation of angular momentum is not guaranteed.
$\nabla \tilde{W}_{ij}$ is not strictly parallel to the relative position vector in the range of rounding error, therefore the new formulation does not guarantee the angular momentum conservation.
However, we find that the errors are smaller and the angular momentum is better conserved than in the previous formulation(Figure \ref{fig:AngMom}).

\subsection{Conservation of energy}

\begin{figure}
    \begin{center}
        \includegraphics[width=80mm]{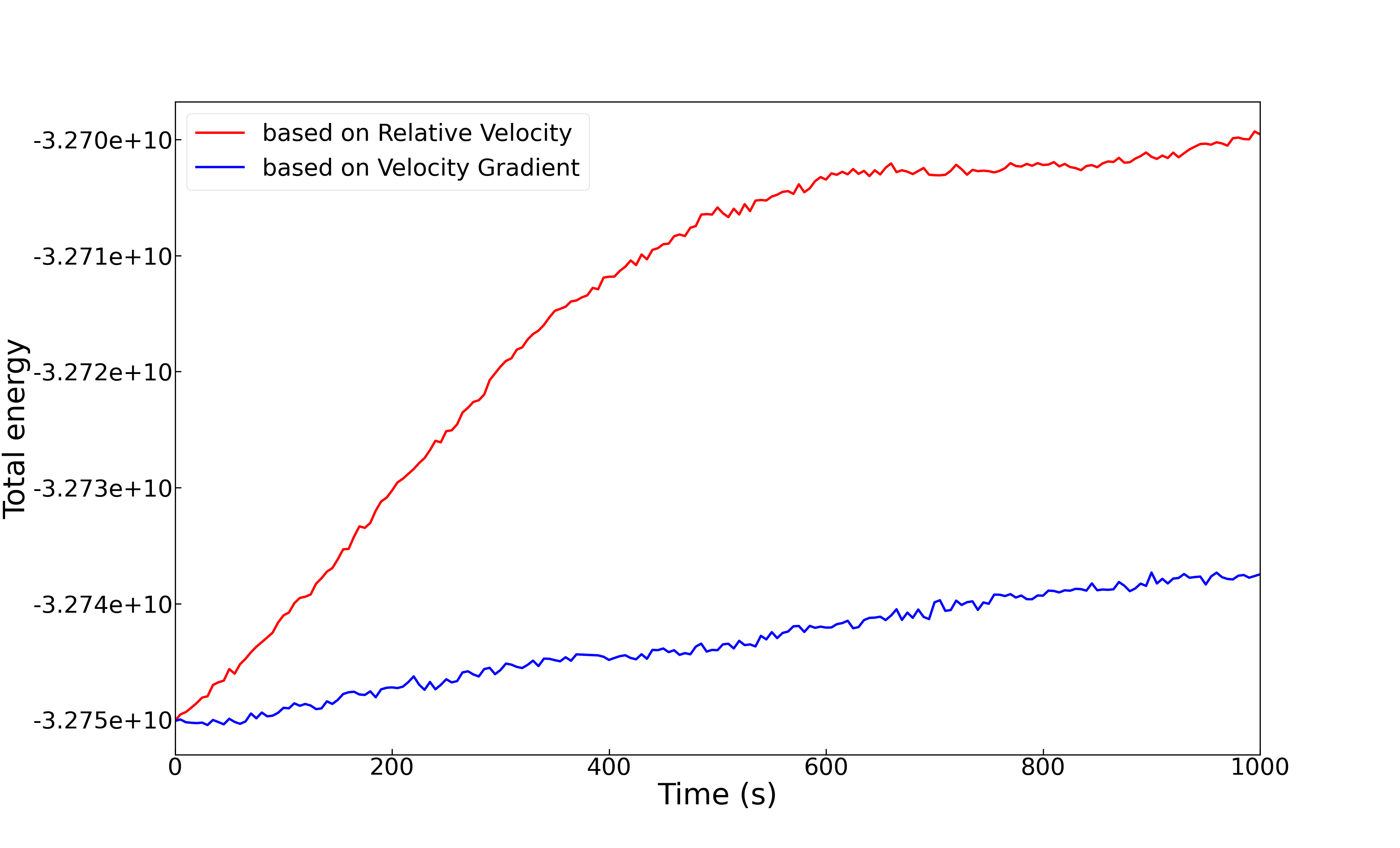}
    \end{center}
    \caption{The time evolution of the total energy (internal energy + mechanical energy) in the numerical test (section \ref{sec:results}) for the formulation based on the relative velocity (red) and the velocity gradient (blue).}
    \label{fig:Eng}
\end{figure}

The SPH method is not written in a conservative form, and conservation of energy is not guaranteed using either the traditional conventional or new formulation of viscosity.
In the DISPH method used in this study, a $\nabla h$ term \citep{springel2002} is introduced to improve the conservation of energy with respect to the pressure term.
The $\nabla h$ term is not introduced for the viscous term because it is difficult to solve the Euler-Lagrange equation for viscous fluids analytically under constraint conditions.
However, as the figure \ref{fig:Eng} shows, the new formulation improves the conservation of energy.
This is likely due to the improved collinearity of the viscous forces in the new viscosity formulation as we saw in \ref{sec:cons_momemtum}, which results in smaller energy errors.

\subsection{The thermodynamical consistency with respect to second law}

\begin{figure}
    \begin{center}
        \includegraphics[width=80mm]{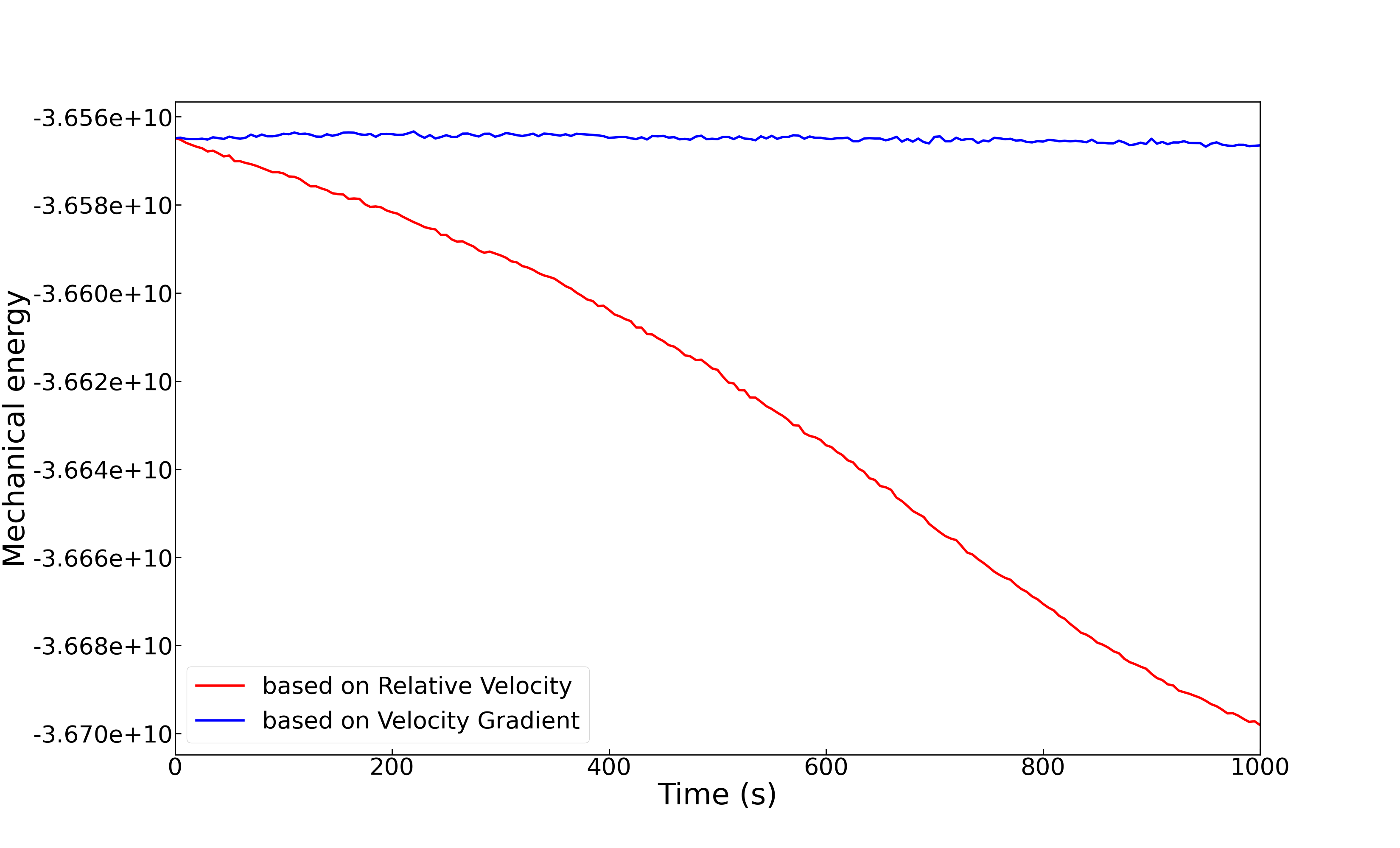}
    \end{center}
    \caption{The time evolution of the mechanical energy in the numerical test (section \ref{sec:results}) for the formulation based on the relative velocity (red) and the velocity gradient (blue).}
    \label{fig:mechanical}
\end{figure}

When dealing with governing equations with dissipative terms, it is required to satisfy the second law of thermodynamics.
However, as mentioned above, the SPH method is not written in a conservative form, and therefore the second law of thermodynamics and other physical laws are only approximately satisfied.
If the second law of thermodynamics is to be satisfied, the mechanical energy must monotonically decrease, but in the case of calculations using the new formulation, the mechanical energy slightly increases in the initial stage of the calculation (Figure \ref{fig:mechanical}).
Several improved SPH methods have been proposed to be thermodynamically consistent \citep[e.g.][]{violeau2012fluid, zhou2023thermodynamically, pavelka2024}, and we plan to adapt these improved methods to our method when simulating the thermal evolution of actual ice satellites in the future.

\bibliographystyle{elsarticle-harv}
\bibliography{reference}

\begin{thebibliography}{42}
\expandafter\ifx\csname natexlab\endcsname\relax\def\natexlab#1{#1}\fi
\providecommand{\url}[1]{\texttt{#1}}
\providecommand{\href}[2]{#2}
\providecommand{\path}[1]{#1}
\providecommand{\DOIprefix}{doi:}
\providecommand{\ArXivprefix}{arXiv:}
\providecommand{\URLprefix}{URL: }
\providecommand{\Pubmedprefix}{pmid:}
\providecommand{\doi}[1]{\href{http://dx.doi.org/#1}{\path{#1}}}
\providecommand{\Pubmed}[1]{\href{pmid:#1}{\path{#1}}}
\providecommand{\bibinfo}[2]{#2}
\ifx\xfnm\relax \def\xfnm[#1]{\unskip,\space#1}\fi
\bibitem[{Agertz et~al.(2007)Agertz, Moore, Stadel, Potter, Miniati, Read, Mayer, Gawryszczak, Kravtsov, Nordlund et~al.}]{agertz2007fundamental}
\bibinfo{author}{Agertz, O.}, \bibinfo{author}{Moore, B.}, \bibinfo{author}{Stadel, J.}, \bibinfo{author}{Potter, D.}, \bibinfo{author}{Miniati, F.}, \bibinfo{author}{Read, J.}, \bibinfo{author}{Mayer, L.}, \bibinfo{author}{Gawryszczak, A.}, \bibinfo{author}{Kravtsov, A.}, \bibinfo{author}{Nordlund, {\AA}.}, et~al., \bibinfo{year}{2007}.
\newblock \bibinfo{title}{Fundamental differences between sph and grid methods}.
\newblock \bibinfo{journal}{Monthly Notices of the Royal Astronomical Society} \bibinfo{volume}{380}, \bibinfo{pages}{963--978}.
\bibitem[{Andersson and Inaba(2005)}]{andersson2005}
\bibinfo{author}{Andersson, O.}, \bibinfo{author}{Inaba, A.}, \bibinfo{year}{2005}.
\newblock \bibinfo{title}{Thermal conductivity of crystalline and amorphous ices and its implications on amorphization and glassy water}.
\newblock \bibinfo{journal}{Phys. Chem. Chem. Phys.} \bibinfo{volume}{7}, \bibinfo{pages}{1441--1449}.
\bibitem[{Ashkenazy et~al.(2018)Ashkenazy, Sayag and Tziperman}]{ashkenazy2018}
\bibinfo{author}{Ashkenazy, Y.}, \bibinfo{author}{Sayag, R.}, \bibinfo{author}{Tziperman, E.}, \bibinfo{year}{2018}.
\newblock \bibinfo{title}{Dynamics of the global meridional ice flow of {{Europa}}'s icy shell}.
\newblock \bibinfo{journal}{Nature Astron.} \bibinfo{volume}{2}, \bibinfo{pages}{43--49}.
\newblock \DOIprefix\doi{10.1038/s41550-017-0326-7}.
\bibitem[{Balsara(1995)}]{balsara1995}
\bibinfo{author}{Balsara, D.S.}, \bibinfo{year}{1995}.
\newblock \bibinfo{title}{von neumann stability analysis of smoothed particle hydrodynamics—suggestions for optimal algorithms}.
\newblock \bibinfo{journal}{J. Comput. Phys.} \bibinfo{volume}{121}, \bibinfo{pages}{357--372}.
\newblock \DOIprefix\doi{https://doi.org/10.1016/S0021-9991(95)90221-X}.
\bibitem[{Cleary and Monaghan(1999)}]{cleary1999}
\bibinfo{author}{Cleary, P.W.}, \bibinfo{author}{Monaghan, J.J.}, \bibinfo{year}{1999}.
\newblock \bibinfo{title}{Conduction modelling using smoothed particle hydrodynamics}.
\newblock \bibinfo{journal}{J. Comput. Phys.} \bibinfo{volume}{148}, \bibinfo{pages}{227--264}.
\newblock \DOIprefix\doi{https://doi.org/10.1006/jcph.1998.6118}.
\bibitem[{Dehnen and Aly(2012)}]{dehnen2012}
\bibinfo{author}{Dehnen, W.}, \bibinfo{author}{Aly, H.}, \bibinfo{year}{2012}.
\newblock \bibinfo{title}{Improving convergence in smoothed particle hydrodynamics simulations without pairing instability}.
\newblock \bibinfo{journal}{MNRAS} \bibinfo{volume}{425}, \bibinfo{pages}{1068--1082}.
\newblock \DOIprefix\doi{10.1111/j.1365-2966.2012.21439.x}.
\bibitem[{Gingold and Monaghan(1977)}]{gingold1977}
\bibinfo{author}{Gingold, R.}, \bibinfo{author}{Monaghan, J.}, \bibinfo{year}{1977}.
\newblock \bibinfo{title}{Smoothed particle hdrodynamics: Theory and application to non-spherical stars}.
\newblock \bibinfo{journal}{MNRAS} \bibinfo{volume}{181}, \bibinfo{pages}{375--389}.
\newblock \DOIprefix\doi{10.1093/mnras/181.3.375}.
\bibitem[{Haldemann et~al.(2020)Haldemann, Alibert, Mordasini and Benz}]{haldemann2020a}
\bibinfo{author}{Haldemann, J.}, \bibinfo{author}{Alibert, Y.}, \bibinfo{author}{Mordasini, C.}, \bibinfo{author}{Benz, W.}, \bibinfo{year}{2020}.
\newblock \bibinfo{title}{{{AQUA}}: {{A Collection}} of {{H}}\$\_2\${{O Equations}} of {{State}} for {{Planetary Models}}}.
\newblock \bibinfo{journal}{A\&A} \bibinfo{volume}{643}, \bibinfo{pages}{A105}.
\newblock \DOIprefix\doi{10.1051/0004-6361/202038367}, \href{http://arxiv.org/abs/2009.10098}{{\tt arXiv:2009.10098}}.
\bibitem[{Hill(1962)}]{hill1962}
\bibinfo{author}{Hill, M.N.}, \bibinfo{year}{1962}.
\newblock \bibinfo{title}{The Sea : ideas and observations on progress in the study of the seas}.
\newblock \bibinfo{publisher}{New York: Interscience}.
\bibitem[{Hopkins(2013)}]{hopkins2013}
\bibinfo{author}{Hopkins, P.F.}, \bibinfo{year}{2013}.
\newblock \bibinfo{title}{A general class of {{Lagrangian}} smoothed particle hydrodynamics methods and implications for fluid mixing problems}.
\newblock \bibinfo{journal}{MNRAS} \bibinfo{volume}{428}, \bibinfo{pages}{2840--2856}.
\newblock \DOIprefix\doi{10.1093/mnras/sts210}.
\bibitem[{Hosono et~al.(2013)Hosono, Saitoh and Makino}]{hosono2013a}
\bibinfo{author}{Hosono, N.}, \bibinfo{author}{Saitoh, T.R.}, \bibinfo{author}{Makino, J.}, \bibinfo{year}{2013}.
\newblock \bibinfo{title}{Density-{{Independent Smoothed Particle Hydrodynamics}} for a {{Non-Ideal Equation}} of {{State}}}.
\newblock \bibinfo{journal}{PASJ} \bibinfo{volume}{65}, \bibinfo{pages}{108}.
\newblock \DOIprefix\doi{10.1093/pasj/65.5.108}.
\bibitem[{Hosono et~al.(2016)Hosono, Saitoh and Makino}]{hosono2016b}
\bibinfo{author}{Hosono, N.}, \bibinfo{author}{Saitoh, T.R.}, \bibinfo{author}{Makino, J.}, \bibinfo{year}{2016}.
\newblock \bibinfo{title}{A comparison of sph artificial viscosities and their impact on the keplerian disk}.
\newblock \bibinfo{journal}{ApJS} \bibinfo{volume}{224}, \bibinfo{pages}{32}.
\bibitem[{Inutsuka(1994)}]{inutsuka1994godunov}
\bibinfo{author}{Inutsuka, S.}, \bibinfo{year}{1994}.
\newblock \bibinfo{title}{Godunov-type sph}.
\newblock \bibinfo{journal}{Memorie della Societ{\`a} Astronomia Italiana, Vol. 65, p. 1027} \bibinfo{volume}{65}, \bibinfo{pages}{1027}.
\bibitem[{Inutsuka(2002)}]{inutsuka2002reformulation}
\bibinfo{author}{Inutsuka, S.i.}, \bibinfo{year}{2002}.
\newblock \bibinfo{title}{Reformulation of smoothed particle hydrodynamics with riemann solver}.
\newblock \bibinfo{journal}{Journal of Computational Physics} \bibinfo{volume}{179}, \bibinfo{pages}{238--267}.
\bibitem[{Kirk and Stevenson(1987)}]{kirk198791}
\bibinfo{author}{Kirk, R.}, \bibinfo{author}{Stevenson, D.}, \bibinfo{year}{1987}.
\newblock \bibinfo{title}{Thermal evolution of a differentiated ganymede and implications for surface features}.
\newblock \bibinfo{journal}{Icarus} \bibinfo{volume}{69}, \bibinfo{pages}{91--134}.
\newblock \DOIprefix\doi{https://doi.org/10.1016/0019-1035(87)90009-1}.
\bibitem[{Landshoff(1955)}]{landshoff1955}
\bibinfo{author}{Landshoff, R.}, \bibinfo{year}{1955}.
\newblock \bibinfo{title}{A numerical method for treating fluid flow in the presence of shocks}.
\newblock \bibinfo{publisher}{Los Alamos, NM: Los Alamos National Laboratory}.
\bibitem[{Lucy(1977)}]{lucy1977}
\bibinfo{author}{Lucy, L.B.}, \bibinfo{year}{1977}.
\newblock \bibinfo{title}{A numerical approach to the testing of the fission hypothesis}.
\newblock \bibinfo{journal}{AJ} \bibinfo{volume}{82}, \bibinfo{pages}{1013}.
\newblock \DOIprefix\doi{10.1086/112164}.
\bibitem[{McNally et~al.(2012)McNally, Lyra and Passy}]{mcnally2012}
\bibinfo{author}{McNally, C.P.}, \bibinfo{author}{Lyra, W.}, \bibinfo{author}{Passy, J.C.}, \bibinfo{year}{2012}.
\newblock \bibinfo{title}{A well-posed kelvin--helmholtz instability test and comparison}.
\newblock \bibinfo{journal}{The Astrophysical Journal Supplement Series} \bibinfo{volume}{201}, \bibinfo{pages}{18}.
\bibitem[{Melosh(1989)}]{melosh1989}
\bibinfo{author}{Melosh, H.J.}, \bibinfo{year}{1989}.
\newblock \bibinfo{title}{Impact Cratering : A Geologic Process}.
\newblock \bibinfo{publisher}{New York: Oxford University Press}.
\bibitem[{Morris et~al.(1997)Morris, Fox and Zhu}]{morris1997}
\bibinfo{author}{Morris, J.P.}, \bibinfo{author}{Fox, P.J.}, \bibinfo{author}{Zhu, Y.}, \bibinfo{year}{1997}.
\newblock \bibinfo{title}{Modeling {{Low Reynolds Number Incompressible Flows Using SPH}}}.
\newblock \bibinfo{journal}{J. Comput. Phys.} \bibinfo{volume}{136}, \bibinfo{pages}{214--226}.
\newblock \DOIprefix\doi{10.1006/jcph.1997.5776}.
\bibitem[{von Neumann and Richtmyer(1950)}]{vonneumann1950}
\bibinfo{author}{von Neumann, J.}, \bibinfo{author}{Richtmyer, R.D.}, \bibinfo{year}{1950}.
\newblock \bibinfo{title}{A method for the numerical calculation of hydrodynamic shocks}.
\newblock \bibinfo{journal}{J. applied Phys} \bibinfo{volume}{21}, \bibinfo{pages}{232--237}.
\bibitem[{Parshikov et~al.(2000)Parshikov, Medin, Loukashenko and Milekhin}]{parshikov2000improvements}
\bibinfo{author}{Parshikov, A.N.}, \bibinfo{author}{Medin, S.A.}, \bibinfo{author}{Loukashenko, I.I.}, \bibinfo{author}{Milekhin, V.A.}, \bibinfo{year}{2000}.
\newblock \bibinfo{title}{Improvements in sph method by means of interparticle contact algorithm and analysis of perforation tests at moderate projectile velocities}.
\newblock \bibinfo{journal}{International Journal of Impact Engineering} \bibinfo{volume}{24}, \bibinfo{pages}{779--796}.
\bibitem[{Pavelka et~al.(2024)Pavelka, Klika and Kincl}]{pavelka2024}
\bibinfo{author}{Pavelka, M.}, \bibinfo{author}{Klika, V.}, \bibinfo{author}{Kincl, O.}, \bibinfo{year}{2024}.
\newblock \bibinfo{title}{Approaches to conservative smoothed particle hydrodynamics with entropy}.
\newblock \URLprefix \url{https://arxiv.org/abs/2406.14229}, \href{http://arxiv.org/abs/2406.14229}{{\tt arXiv:2406.14229}}.
\bibitem[{Pearl et~al.(2022)Pearl, Raskin and Owen}]{pearl2022fsisph}
\bibinfo{author}{Pearl, J.M.}, \bibinfo{author}{Raskin, C.D.}, \bibinfo{author}{Owen, J.M.}, \bibinfo{year}{2022}.
\newblock \bibinfo{title}{Fsisph: An sph formulation for impacts between dissimilar materials}.
\newblock \bibinfo{journal}{Journal of Computational Physics} \bibinfo{volume}{469}, \bibinfo{pages}{111533}.
\bibitem[{Porco et~al.(2006)Porco, Helfenstein, Thomas, Ingersoll, Wisdom, West, Neukum, Denk, Wagner, Roatsch et~al.}]{porco2006}
\bibinfo{author}{Porco, C.C.}, \bibinfo{author}{Helfenstein, P.}, \bibinfo{author}{Thomas, P.}, \bibinfo{author}{Ingersoll, A.}, \bibinfo{author}{Wisdom, J.}, \bibinfo{author}{West, R.}, \bibinfo{author}{Neukum, G.}, \bibinfo{author}{Denk, T.}, \bibinfo{author}{Wagner, R.}, \bibinfo{author}{Roatsch, T.}, et~al., \bibinfo{year}{2006}.
\newblock \bibinfo{title}{Cassini observes the active south pole of enceladus}.
\newblock \bibinfo{journal}{Science} \bibinfo{volume}{311}, \bibinfo{pages}{1393--1401}.
\bibitem[{Price(2008)}]{price2008modelling}
\bibinfo{author}{Price, D.J.}, \bibinfo{year}{2008}.
\newblock \bibinfo{title}{Modelling discontinuities and kelvin--helmholtz instabilities in sph}.
\newblock \bibinfo{journal}{Journal of Computational Physics} \bibinfo{volume}{227}, \bibinfo{pages}{10040--10057}.
\bibitem[{Price et~al.(2018)Price, Wurster, Tricco, Nixon, Toupin, Pettitt, Chan, Mentiplay, Laibe, Glover et~al.}]{price2018phantom}
\bibinfo{author}{Price, D.J.}, \bibinfo{author}{Wurster, J.}, \bibinfo{author}{Tricco, T.S.}, \bibinfo{author}{Nixon, C.}, \bibinfo{author}{Toupin, S.}, \bibinfo{author}{Pettitt, A.}, \bibinfo{author}{Chan, C.}, \bibinfo{author}{Mentiplay, D.}, \bibinfo{author}{Laibe, G.}, \bibinfo{author}{Glover, S.}, et~al., \bibinfo{year}{2018}.
\newblock \bibinfo{title}{Phantom: A smoothed particle hydrodynamics and magnetohydrodynamics code for astrophysics}.
\newblock \bibinfo{journal}{Publications of the Astronomical Society of Australia} \bibinfo{volume}{35}, \bibinfo{pages}{e031}.
\bibitem[{Reese et~al.(1999)Reese, Solomatov and Moresi}]{reese1999a}
\bibinfo{author}{Reese, C.}, \bibinfo{author}{Solomatov, V.}, \bibinfo{author}{Moresi, L.N.}, \bibinfo{year}{1999}.
\newblock \bibinfo{title}{Non-{{Newtonian Stagnant Lid Convection}} and {{Magmatic Resur}} facing on {{Venus}}}.
\newblock \bibinfo{journal}{Icarus} \bibinfo{volume}{139}, \bibinfo{pages}{67--80}.
\newblock \DOIprefix\doi{10.1006/icar.1999.6088}.
\bibitem[{Richtmyer(1948)}]{richtmyer1948}
\bibinfo{author}{Richtmyer, R.D.}, \bibinfo{year}{1948}.
\newblock \bibinfo{title}{Proposed numerical method for calculation of shocks}.
\newblock \bibinfo{publisher}{Los Alamos, NM: Los Alamos National Laboratory}.
\bibitem[{Saitoh and Makino(2013)}]{saitoh2013}
\bibinfo{author}{Saitoh, T.R.}, \bibinfo{author}{Makino, J.}, \bibinfo{year}{2013}.
\newblock \bibinfo{title}{A {{DENSITY-INDEPENDENT FORMULATION OF SMOOTHED PARTICLE HYDRODYNAMICS}}}.
\newblock \bibinfo{journal}{AJ} \bibinfo{volume}{768}, \bibinfo{pages}{44}.
\newblock \DOIprefix\doi{10.1088/0004-637X/768/1/44}.
\bibitem[{{Sijacki} and {Springel}(2006)}]{sijacki2006}
\bibinfo{author}{{Sijacki}, D.}, \bibinfo{author}{{Springel}, V.}, \bibinfo{year}{2006}.
\newblock \bibinfo{title}{{Hydrodynamical simulations of cluster formation with central AGN heating}}.
\newblock \bibinfo{journal}{MNRAS} \bibinfo{volume}{366}, \bibinfo{pages}{397--416}.
\newblock \DOIprefix\doi{10.1111/j.1365-2966.2005.09860.x}, \href{http://arxiv.org/abs/astro-ph/0509506}{{\tt arXiv:astro-ph/0509506}}.
\bibitem[{Sparks et~al.(2016)Sparks, Hand, McGrath, Bergeron, Cracraft and Deustua}]{sparks2016}
\bibinfo{author}{Sparks, W.B.}, \bibinfo{author}{Hand, K.}, \bibinfo{author}{McGrath, M.}, \bibinfo{author}{Bergeron, E.}, \bibinfo{author}{Cracraft, M.}, \bibinfo{author}{Deustua, S.}, \bibinfo{year}{2016}.
\newblock \bibinfo{title}{Probing for evidence of plumes on europa with hst/stis}.
\newblock \bibinfo{journal}{AJ} \bibinfo{volume}{829}, \bibinfo{pages}{121}.
\bibitem[{Springel and Hernquist(2002)}]{springel2002}
\bibinfo{author}{Springel, V.}, \bibinfo{author}{Hernquist, L.}, \bibinfo{year}{2002}.
\newblock \bibinfo{title}{Cosmological smoothed particle hydrodynamics simulations: the entropy equation}.
\newblock \bibinfo{journal}{Monthly Notices of the Royal Astronomical Society} \bibinfo{volume}{333}, \bibinfo{pages}{649--664}.
\bibitem[{Takeyama et~al.(2017)Takeyama, Saitoh and Makino}]{takeyama2017}
\bibinfo{author}{Takeyama, K.}, \bibinfo{author}{Saitoh, T.R.}, \bibinfo{author}{Makino, J.}, \bibinfo{year}{2017}.
\newblock \bibinfo{title}{Variable inertia method: {{A}} novel numerical method for mantle convection simulation}.
\newblock \bibinfo{journal}{New Astron.} \bibinfo{volume}{50}, \bibinfo{pages}{82--103}.
\newblock \DOIprefix\doi{10.1016/j.newast.2016.07.002}.
\bibitem[{Tillotson(1962)}]{tillotson1962}
\bibinfo{author}{Tillotson, J.H.}, \bibinfo{year}{1962}.
\newblock \bibinfo{title}{Metallic {{Equations}} of {{State For Hypervelocity Impact}}}.
\newblock \bibinfo{publisher}{San Diego: General Atomic Report}.
\bibitem[{Tobie(2003)}]{tobie2003}
\bibinfo{author}{Tobie, G.}, \bibinfo{year}{2003}.
\newblock \bibinfo{title}{Tidally heated convection: {{Constraints}} on {{Europa}}'s ice shell thickness}.
\newblock \bibinfo{journal}{J. Geophy. Res.} \bibinfo{volume}{108}, \bibinfo{pages}{5124}.
\newblock \DOIprefix\doi{10.1029/2003JE002099}.
\bibitem[{Tricco(2019)}]{tricco2019kelvin}
\bibinfo{author}{Tricco, T.S.}, \bibinfo{year}{2019}.
\newblock \bibinfo{title}{The kelvin--helmholtz instability and smoothed particle hydrodynamics}.
\newblock \bibinfo{journal}{Monthly Notices of the Royal Astronomical Society} \bibinfo{volume}{488}, \bibinfo{pages}{5210--5224}.
\bibitem[{Violeau(2012)}]{violeau2012fluid}
\bibinfo{author}{Violeau, D.}, \bibinfo{year}{2012}.
\newblock \bibinfo{title}{Fluid mechanics and the SPH method: theory and applications}.
\newblock \bibinfo{publisher}{Oxford University Press}.
\bibitem[{Wadsley et~al.(2017)Wadsley, Keller and Quinn}]{wadsley2017gasoline2}
\bibinfo{author}{Wadsley, J.W.}, \bibinfo{author}{Keller, B.W.}, \bibinfo{author}{Quinn, T.R.}, \bibinfo{year}{2017}.
\newblock \bibinfo{title}{Gasoline2: a modern smoothed particle hydrodynamics code}.
\newblock \bibinfo{journal}{Monthly Notices of the Royal Astronomical Society} \bibinfo{volume}{471}, \bibinfo{pages}{2357--2369}.
\bibitem[{Wendland(1995)}]{wendland1995}
\bibinfo{author}{Wendland, H.}, \bibinfo{year}{1995}.
\newblock \bibinfo{title}{Piecewise polynomial, positive definite and compactly supported radial functions of minimal degree}.
\newblock \bibinfo{journal}{Adv. Comput. Math.} \bibinfo{volume}{4}, \bibinfo{pages}{389--396}.
\newblock \DOIprefix\doi{10.1007/BF02123482}.
\bibitem[{Yuasa and Mori(2024)}]{yuasa2024}
\bibinfo{author}{Yuasa, T.}, \bibinfo{author}{Mori, M.}, \bibinfo{year}{2024}.
\newblock \bibinfo{title}{Novel hydrodynamic schemes capturing shocks and contact discontinuities and comparison study with existing methods}.
\newblock \bibinfo{journal}{New Astronomy} \bibinfo{volume}{109}, \bibinfo{pages}{102208}.
\newblock \URLprefix \url{http://dx.doi.org/10.1016/j.newast.2024.102208}, \DOIprefix\doi{10.1016/j.newast.2024.102208}.
\bibitem[{Zhou et~al.(2023)Zhou, Zhao, Bi, Zhang, Wang and Li}]{zhou2023thermodynamically}
\bibinfo{author}{Zhou, Z.}, \bibinfo{author}{Zhao, Y.}, \bibinfo{author}{Bi, J.}, \bibinfo{author}{Zhang, Y.}, \bibinfo{author}{Wang, C.}, \bibinfo{author}{Li, Y.}, \bibinfo{year}{2023}.
\newblock \bibinfo{title}{A thermodynamically consistent sph-pfm model for modelling crack propagation and coalescence in rocks}.
\newblock \bibinfo{journal}{Theoretical and Applied Fracture Mechanics} \bibinfo{volume}{127}, \bibinfo{pages}{104085}.

\end{thebibliography}

\end{document}